\documentclass[sigconf]{acmart} 

\AtBeginDocument{%
  }

\copyrightyear{2026}
\acmYear{2026}
\setcopyright{cc}
\setcctype{by}
\acmConference[ICSE '26]{2026 IEEE/ACM 48th International Conference on Software Engineering}{April 12--18, 2026}{Rio de Janeiro, Brazil}
\acmBooktitle{2026 IEEE/ACM 48th International Conference on Software Engineering (ICSE '26), April 12--18, 2026, Rio de Janeiro, Brazil}
\acmPrice{}
\acmDOI{10.1145/3744916.3773165}
\acmISBN{979-8-4007-2025-3/26/04}

\usepackage{subcaption}
\usepackage{siunitx}

\newcommand\code[1]{{\tt \small #1}}

\usepackage{array}
\newcolumntype{L}[1]{>{\raggedright\arraybackslash}p{#1}} 
\newcolumntype{C}[1]{>{\centering\arraybackslash}p{#1}}   
\newcolumntype{R}[1]{>{\raggedleft\arraybackslash}p{#1}}  

\usepackage{listings}

\lstnewenvironment{inlinelisting}{
\lstset{language=Java,
numbers=none,
basicstyle=\fontsize{8}{8}\selectfont\ttfamily
}}{}

\begin{document}

\title{JEDI: Java Evaluation of Declarative and Imperative Queries}
\subtitle{Benchmarking the Java Stream API}

\author{Filippo Schiavio}
\email{filippo.schiavio@usi.ch}
\orcid{0000-0001-9023-0720}
\affiliation{%
  \institution{Università della Svizzera italiana (USI)}
  \city{Lugano}
  \country{Switzerland}
}

\author{Walter Binder}
\email{walter.binder@usi.ch}
\orcid{0000-0002-2477-2182}
\affiliation{%
  \institution{Università della Svizzera italiana (USI)}
  \city{Lugano}
  \country{Switzerland}
}

\begin{abstract}

The Java Stream API aims at increasing developer productivity thanks to an easy-to-read declarative syntax to express computations.
It also simplifies parallel computing, providing a high-level abstraction on top of common parallelization aspects. 
Unfortunately, there is a lack of benchmarks specifically targeting stream-based applications. 
Such a lack of benchmarks makes it difficult for researchers and developers of the Java class library to optimize the Stream API.
Moreover, in the absence of dedicated benchmarks, it is difficult to analyze the performance of streams to suggest developers how to write efficient code using the API.

In this work we present JEDI, a benchmark suite that targets the Stream API. 
JEDI is automatically generated by converting SQL benchmarks into Java benchmarks.
Our code generator supports targets different implementations (both stream-based and imperative) for the same query.
The ultimate goal of our benchmark suite---and the main contribution of this work---is to analyze the performance of the different implementations to spot inefficient code structures and better alternatives, suggesting best practices to Java developers.
Among the multiple implementations we generate, we focus on different parallelization strategies and explain the most efficient parallelization strategies based on characteristics of the processed data.
Finally, the code generation producing imperative code defines of a baseline that can guide researchers and Java implementers to optimize the Stream API.

\end{abstract}



\maketitle

\section{Introduction}

Java~8 has introduced the Stream API~\cite{oracle-stream} in the Java Class Library (JCL) as a declarative way to express computations over collections as well as infinite data sources such as generator functions.
The Stream API offers methods to create streams from any type of collection as well as from arrays, to filter elements and apply data transformations according to user-defined functions---usually expressed as lambda functions---as well as performing aggregations of data elements using reductions.

Such a declarative programming style for processing collections is commonly preferred by developers compared with the traditional imperative programming style of Java (i.e., loop-based code), as it is easier and faster to write, and results in code which is less error-prone and easier to read and maintain~\cite{stream_rct}.
However, from a performance point of view, the Stream API is often slower than equivalent imperative code~\cite{streamliner, parstreamliner}.
Unfortunately, there is no benchmark suite focusing on the Stream API. 
As a result, there is no guideline that would suggest developers best practices that help minimize the overheads introduced by the abstractions of the Stream API.
Moreover, the Java language developers lack benchmarks that would help them optimize the Stream API.

In this work, we present JEDI (Java Evaluation of Declarative and Imperative queries), a benchmark suite for the Java Stream API automatically generated by converting SQL queries, mostly from the established database (DB) benchmark TPC-H~\cite{tpch}, into Java source code.
JEDI is composed of multiple semantically equivalent implementations that make use of different stream operations. 
Among these implementations, we also generate streams that exploit parallelization using different reduction strategies, i.e., merging thread-local results in a fork-join style, lock-based synchronization on a shared state, and lock-free synchronization using thread-safe atomic counters as shared state.

Thanks to this approach, we analyze the performance of the Stream API, comparing alterative operations and programming idioms that allow developers to express the same computational task. 
We also analyze in which case each setting results in better performance.
This evaluation can help application developers to use the Stream API more efficiently.

Finally, in JEDI, we generate imperative implementations (i.e., based on loops over arrays) of the same queries. 
This implementations allows us to evaluate the so-called ``overhead of the Stream API'' by computing the speedup factors of the imperative versions against the stream-based version.
This evaluation can be beneficial to the JCL developers in optimizing the Stream API.
In this way, anyone can easily compare the performance of stream-based implementations against equivalent imperative implementations which, by design (i.e., exploiting code generation) use exactly the same order of operations on the same data structures.

The paper contributions are:
\begin{enumerate}
    \item A performance analysis of different equivalent implementations of relational queries using the Stream API, from which we derive a set of best practices for an efficient usage of the Stream API based on our evaluation, which can serve as a guideline for developers.
    \item A performance comparison of the best sequential stream implementation against an equivalent imperative version (applying the same operations, in the same order and using the same data structure) to evaluate the overhead of the Stream API. 
    \item An evaluation of the performance of parallel streams with different parallel reduction strategies.
    \item We release both JEDI~\cite{jedi}, our generated benchmark suite, and its code generator~\cite{s2s-code} as open-source software.\footnote{A reproducible artifact  is available at \url{http://github.com/usi-dag/JEDI-artifact}}
\end{enumerate}

First, we present background information (Sec.~\ref{sec:background}) and the methodology to convert SQL queries into Java source code using the Stream API (Sec.~\ref{sec:methodology}).
We explain the supported code-generation options (Sec.~\ref{sec:codegen_options}) and introduce JEDI, our benchmark suite generated from the TPC-H DB benchmark (Sec.~\ref{sec:bench}).
We use JEDI to analyze stream performance and to answer the research questions of this work (Sec.~\ref{sec:eval}).
We discuss potential threats to validity (Sec.~\ref{sec:validity}) and related work (Sec.~\ref{sec:relwork}), before concluding (Sec.~\ref{sec:conclusion}).

\section{Background}\label{sec:background}
Here, we introduce background information and the terminology used in this paper.
We first describe the Java Stream API and our automatic approach to generate JEDI with the goal of analyzing stream performance.

\subsection{Java Streams}
A stream is commonly defined as a pipeline of data-processing operators. Each operator is defined by a Java method in the class \code{java.util.stream.Stream}.\footnote{In the same JCL package, there are classes to handle primitive streams, i.e., \code{IntStream}, \code{LongStream}, and \code{DoubleStream}; we omit their description for brevity.}
Streams are evaluated lazily and at most once. 
Specifically, streams are executed when a terminal operation is invoked, i.e., stream creations and intermediate operations do not process any source data. 
Instead, these methods append operations to a data structure that represents the stream pipeline.
When the terminal operation is invoked, the source data elements are processed by all the operations in the pipeline.

Stream methods fall into three main categories: stream-creation methods, intermediate operations, and terminal operations.

\textbf{Stream-creation} methods are defined as methods whose return type is an instance of \code{Stream}. 
Every stream pipeline starts with a stream-creation method.
Stream sources can be arrays or collections, as well as generator functions.

\textbf{Intermediate operations} are methods defined in the \code{Stream} class whose return type is a stream. Common examples of intermediate operations are \code{filter(...)} and \code{map(...)}, which take a function as a parameter and return a stream. The former method returns a stream with the elements of the original stream filtered by the given boolean function. The latter methods return a stream whose elements are obtained by applying the given function (i.e., a transformation) to the elements of the original stream.

Two important intermediate operations in the context of our work are \code{flatMap} and \code{mapMulti}.
Both operations are designed to perform one-to-many stream transformations.
The main difference between them is their parameters. \code{flatMap} accepts a mapper function from the original stream elements into a new stream, which is used to transform a single element into multiple ones that are then flattened into a single stream. The
\code{mapMulti} operation accepts a \code{BiConsumer}~\cite{jdk23biconsumer}, i.e., a function with two arguments. 
The first argument of this function is an element of the original stream, the second one is a \code{Consumer}~\cite{jdk23consumer} (i.e., a function with a single argument), which can be called multiple times after transforming the stream element into multiple ones.

\textbf{Terminal operations} are methods defined in the \code{Stream} class whose return type is not a stream. 
These operations are commonly used to express aggregations.
There are two main operations in this category: reductions (expressed with the \code{reduce(...)} method) and collectors (expressed with the \code{collect(...)} method). 
The former are well known in data-flow languages, while the latter are peculiar in the context of object-oriented programming. 
In particular, a collector is a special case of a reduction which accumulates partial results in result containers, in contrast to a reduction that accumulates partial results applying a binary operator in a purely functional way.

\textbf{Order of execution} may enforce processing the elements in the stream respecting the \emph{encounter order} of the elements. 
Streams can be ordered or unordered, depending on the stream source and intermediate operations.
Some streams sources are inherently ordered (e.g., lists and arrays), others are not (e.g., sets). 
The intermediate operations \code{sorted(...)} enforces streams to be ordered, while \code{unordered(...)} removes the ordering constraint.

If a stream is ordered, most operations are constrained to operate on the elements in their  \emph{encountered order}.
In principle, unordered streams can be implemented more efficiently than ordered ones, since there is no constraint of respecting an encounter order of the stream elements~\cite{jdk23streampackage_order,khatchadourian-opt-stream-4,parstreamliner}

\textbf{Parallel streams} are executed leveraging parallel computation.
Developers can exploit parallel execution simply by marking a stream as parallel, by invoking the intermediate operation \code{parallel()}.
Without any other user intervention, parallel streams are executed in a classic fork/join programming style, i.e., the stream source is partitioned into data chunks, and encapsulated into tasks, splitting larger tasks into smaller ones. 
Task execution can run in parallel and each thread works on thread-local partial results. 
When two forked tasks have been computed, their local results can be joined (i.e., merged).
Additionally, the terminal operations can be marked as \code{concurrent} by the developer.
In this case, the parallel execution does not follow the typical join phase merging thread-local results; instead, all threads work on a thread-safe shared data structure that contains the final result.
Concurrent execution has the benefit that it does not require a merging phase, but may introduce high contention due to the synchronization operations implemented in the thread-safe result container, which are needed to prevent races when accessing the shared state.

\section{Methodology}\label{sec:methodology}
To create Java benchmarks that stress the Stream API, we propose to automatically convert SQL queries into Java source code.
We convert SQL queries because one of the most interesting uses of streams is data processing~\cite{java8inaction} and SQL is a declarative language designed for this purpose. 
Particularly, streams are mostly used for processing in-memory collections; likewise, many SQL databases, particularly those designed for analytical workloads, are in-memory databases~\cite{in-memory-db,zhang2015memory} (i.e., the SQL tables are stored in main memory).

Another important motivation is that there are many established benchmarks carefully designed by data-processing experts which are composed of queries covering complementary features of data processing. 
Thus, we believe that translating such SQL queries yields relevant stream workloads. 
Finally, we note that since SQL is a language, it can be automatically converted into different implementations (e.g., code using the Stream API or loop-based imperative code, as we done in this work) using established SQL query-compilation techniques~\cite{hyper,lb2}. 
Besides simplifying the automatic benchmark generation, the use of SQL makes it easier to test the generated implementations, since their results can be compared with the results of the queries executed in any SQL database.

Our conversion first generates Java classes representing the SQL schema as done in Object-Relational Mapping frameworks.
Each SQL table is implemented as an array of objects of the class representing the table, resulting in an in-memory DB. 
Then, it converts a SQL query to Java code composed of multiple stream pipelines.
Pipelines created from table scans create streams from arrays using the stream creation method \code{Arrays.stream(...)}.
Projections are converted to \code{map(...)} operations, predicates to \code{filter(...)} operations. 
The \code{LIMIT} operator is converted to the equivalent \code{limit(...)} operation, and the \code{ORDER BY} operator is converted to the \code{sort(...)} operation.
Aggregations (both grouped, i.e., those using \code{GROUP BY} and non-grouped ones) are converted into the terminal operator \code{collect(...)} using a generated class as a collector to perform the aggregation.
Grouped aggregations make use of the JCL collector \code{Collectors.groupingBy(...)} before using the automatically generated collector.
Finally, joins are converted into two pipelines. As commonly done by existing query engines~\cite{umbra,hyper,dynq,dynq_journal} the left side of the join is the \emph{build} side (i.e., it puts the elements of the left side into a dedicated in-memory data structure), while the right side is the \emph{probe} side (i.e., it accesses the data structure built during the execution of the left side for each element of the right side). 

In this work, we extend S2S~\cite{s2s}, an existing tool that converts a (\code{SELECT}) SQL query into equivalent Java source code that makes use of the Stream API.
First, we extend S2S by implementing the missing SQL operations that are required to convert TPC-H, e.g., left-joins, semi-joins, and anti-joins.
More importantly, we extend S2S to generate multiple equivalent implementations given a single SQL query, as described in Sec.~\ref{sec:codegen_options}.

Figure~\ref{fig:sql:example} shows a SQL query and Figure~\ref{fig:sql:example:stream} shows the code generated by S2S. 
The \code{exec} method (Figure~\ref{fig:sql:example:stream}, lines~\ref{lines:ex:exec-begin}--\ref{lines:ex:exec-end}) executes the query.
The generated code first creates a stream from the input array (line~\ref{lines:ex:streamcreation1}) and then creates a hash-map for the \code{customer} table, grouped by \code{id} (line~\ref{lines:ex:tomap}).
This is the left side of the join between orders and customers.
Then, it creates a stream for \code{order} (line~\ref{lines:ex:streamcreation2}), and executes the predicates (lines~\ref{lines:ex:filter1a}, \ref{lines:ex:filter1b}, \ref{lines:ex:filter2}).
Then, it sorts the elements in the stream by the field \code{amount} (line~\ref{lines:ex:sorted}) using the comparator defined in the field \code{comp} (line~\ref{lines:ex:comp}), taking only the first 100 rows (line~\ref{lines:ex:limit}).
Then, it uses a \code{map} operation to evaluate the expression \code{amount * discount} (line~\ref{lines:ex:map2}).
The implementation then performs a join operation between the \code{orders} and \code{customer} tables. 
To this end, it uses a \code{flatMap} stream operation by getting, for each element in \code{order}, the related rows in \code{customer} using the previously created hash-map (lines~\ref{lines:ex:flatmap}--\ref{lines:ex:flatmapend}).
Finally, the generated code performs the group-by operation using the \code{groupingBy} method (line~\ref{lines:ex:aggcallbegin} and the aggregation (\code{SUM(amount))} using a collector (\code{Agg}, lines~\ref{lines:ex:aggbegin}--\ref{lines:ex:aggend}).
The final result is returned in the form of a list created from the values in the hash-map built in the described aggregation (lines~\ref{lines:ex:returnbegin}--\ref{lines:ex:returnend}).

\begin{figure}
\begin{lstlisting}[numbers=none]
WITH high_value_purchases AS (
    SELECT * FROM orders
    WHERE EXTRACT(YEAR FROM odate) >= 2024
    AND shipcountry = "Brasil"
    ORDER BY amount DESC
    LIMIT 100
)

SELECT C.name, SUM(P.amount * P.discount)
FROM customer C, high_value_purchases P
WHERE C.id = P.custumer_id
GROUP BY C.name;
\end{lstlisting}
\caption{SQL query as running example.}
\label{fig:sql:example}\vspace{-2mm}
\end{figure}

\begin{figure}
\begin{lstlisting}[firstline=10,escapechar=|]
public class Query {
	Comparator<P0> comp = Comparator
		.comparing((P0 p) -> p.amount())
		.reversed();  |\label{lines:ex:comp}|
	
	record J0(...) {} // fields omitted
	record P0(...) {} // fields omitted
	
	class Agg {  |\label{lines:ex:aggbegin}|
		String custumerName; int tot;
		void acc(J0 row) {
			customerName = row.customer_name; 
			tot += row.price_discount; 
		}
		Agg combine(Agg other) {
			tot += other.tot;
			return this;
		}
		static Collector<J0, Agg, Agg> collector(){
			return Collectors.of(
				Agg::new, Agg::acc, Agg::combine);
		}
	} |\label{lines:ex:aggend}|
	public List<Agg> exec(DB db) {|\label{lines:ex:exec-begin}|
		var hm = Arrays.stream(db.customer) |\label{lines:ex:streamcreation1}|
			.collect(groupingBy(row -> row.id)); |\label{lines:ex:tomap}|
		List<P0> E = Collections.emptyList();
		return Arrays.stream(db.orders) |\label{lines:ex:streamcreation2}|
			.filter(row -> |\label{lines:ex:filter1a}|
					row.odate.getYear() >= 2024))|\label{lines:ex:filter1b}|
			.filter(row -> row.shipcountry.equals("Brasil")))|\label{lines:ex:filter2}|
			.sorted(comp) |\label{lines:ex:sorted}|
			.limit(100) |\label{lines:ex:limit}|
			.map(row -> new P0(row.customer_id, row.amount * row.discount)) |\label{lines:ex:map2}|
			.flatMap(r -> |\label{lines:ex:flatmap}|
				hm.getOrDefault(r.custumer_id, E)
					.stream()
					.map(l -> new J0(...)))  |\label{lines:ex:flatmapend}|
			.collect(Collectors.groupingBy(|\label{lines:ex:aggcallbegin}|
				row -> row.customer_name,
				HashMap::new,
				Agg.collector()))|\label{lines:ex:aggcallend}|
			.values()|\label{lines:ex:returnbegin}|
			.stream().toList();|\label{lines:ex:returnend}|
	}|\label{lines:ex:exec-end}|
}
\end{lstlisting}
\caption{Code generated for the example query (Figure~\ref{fig:sql:example}).}
\label{fig:sql:example:stream}
\end{figure}

Each generated implementation is automatically tested with JUnit~\cite{junit} tests, comparing its results with the original query executed in a DB (our implementation currently uses DuckDB~\cite{duckdb}).
\section{Code generation options}\label{sec:codegen_options}
In this section, we describe our S2S extensions to generate multiple implementation alternatives for a given query. 

{\bf O1) Filter fusion.}
SQL queries often involve filtering data elements using a conjunction of predicates.
As an example, the SQL query in Figure~\ref{fig:sql:example} filters the rows in the \code{order} table with two predicates, i.e., the orders requested (1)~in a year from 2024, (2)~shipped in Brasil.
As shown in Figure~\ref{fig:sql:example:stream}, given a conjunction of predicates, S2S generates a \code{Stream.filter(...)} invocation for each predicate.

Filter fusion, indicates that a conjunction of predicates should use a single call to \code{Stream.filter(...)}, combining multiple predicates into a single boolean lambda expression that uses the \code{\&\&} operator.
As an example, when this option is enabled in our code generator, the code for implementing the predicate conjunction in the example query (Figure~\ref{fig:sql:example:stream}, lines~\ref{lines:ex:filter1a},~\ref{lines:ex:filter1b},~\ref{lines:ex:filter2}) would be the following:

\begin{lstlisting}[numbers=none]
.filter(row -> row.odate.getYear() >= 2024 && row.shipcountry.equals("Brasil")))
\end{lstlisting}

This option is designed to explore if it is more efficient to use multiple calls to \code{Stream.filter(...)} with simpler lambda expressions, or a single call to \code{Stream.filter(...)} with a more complex lambda expression.

{\bf O2) Join implementation.}
We also explore an alternative strategy to implement the SQL join operator.
As Figure~\ref{fig:sql:example:stream} shows, the proposed conversion from SQL to stream implements the join by collecting elements from the build side of the join in a data-structure (i.e., a hash-map in case of hash joins and a list in case of nested loop joins).
Then, it uses the \code{Stream.flatMap(...)} method to iterate over the matching pairs of rows.%
\footnote{If the SQL schema makes explicit use of foreign keys, DBs guarantee that there can be only a single matching pair. However, S2S does not implement this feature and, conservatively, it stores all rows with the same key in a list. 
This approach may result in lists with a single element, ensuring correct execution also when the joined tables do not make use of foreign keys.} 
This code-generation option indicates that joins should use \code{Stream.mapMulti(...)} instead of \code{Stream.flatMap(...)} to implement the join operator, as in the example query (Figure~\ref{fig:sql:example:stream}, lines~\ref{lines:ex:flatmap}--\ref{lines:ex:flatmapend}).

{\bf O3) Parallel evaluation.}
The original S2S tool could generate only sequential streams.
This code-generation option explores multiple parallel implementation strategies, namely:
\begin{enumerate}
    \item (P) A parallel stream is generated with calls to \code{.parallel()}. 
    Parallel streams implement joins and group-by operators with the \code{groupingBy} operation.
    Internally, the implementation uses thread-local maps\footnote{Instances of the class \code{java.util.HashMap}, which is not thread-safe.}.
    Particularly, this operation needs to merge maps when combining the results of forked sub-tasks. 
    These parallel streams are ordered, i.e., they have to process the elements from the stream source in encounter order. 
    
    \item (PU) An unordered parallel stream 
    is generated with calls to \code{.parallel().unordered()}.
    PU streams process data elements as done with P streams, with the difference that there is no constraint on processing elements in encounter order. 
    
    \item (CG) An PU stream that uses the JCL concurrent grouping-by operation, i.e., it calls the method \code{groupingByConcurrent} instead of \code{groupingBy}.  
    This method uses a single thread-safe map\footnote{An instance of the class \code{java.util.concurrent.ConcurrentHashMap}.} instance shared among multiple threads, so there is no merging phase, in contrast to PU.
    Remarkably, similarly to PU, the downstream collector of the \code{groupingByConcurrent} operation (i.e., the collector generated by S2S that implements the group-by or join operator) is not thread-safe.
    When the downstream collectors are not marked as thread-safe, the JCL internally makes use of (intrinsic) lock-based synchronization when accessing the state of the values in the map.
    
    \item (CGCC) A CG stream that also uses thread-safe (concurrent) downstream collectors. In contrast to CG, a CGCC stream does not require lock-based synchronization.
    Instead, it exploits atomic updates using compare-and-swap\footnote{Atomic operations are provided in the package \code{java.util.concurrent.atomic}.} operations in the collector created by our code generator. 
    
\end{enumerate}

\section{JEDI: The Generated Benchmark Suite}\label{sec:bench}

Our methodology (Sec~\ref{sec:methodology}) is general and can be applied to arbitrary SQL queries.
In this work, we generate JEDI from TPC-H~\cite{tpch}, a widely used DB benchmark.
We have selected TPC-H because it is the most used data-analytics benchmarks in the DB community, cited by more than thousand research papers~\cite{chokepoints2}, and covers many complementary features of data-processing~\cite{chokepoints1}.
TPC-H comes with a data generator that is parametric on the scale factor (SF), which determines the final dataset size. 
SF set to $N$ yields a dataset of $N$~GB in UTF encoding.
We generate JEDI by converting each TPC-H query into multiple, semantically equivalent, implementations. 
In this way, we can analyze the performance characteristic of different usage patterns of the Stream API.

As discussed in the literature~\cite{chokepoints1,chokepoints2,everything-vect-comp}, most performance challenges of the TPC-H benchmark are present in five queries, i.e., Q01, Q03, Q06, Q09, and Q18. 
Q01 and Q06 are the only two queries on a single table, i.e., they do not make use of the join operator. 
Q01 performs a low-cardinality grouped aggregation (i.e., there are only 4 distinct groups), and executes many arithmetic operations (sum and averages) for each aggregated group.
Q06 is dominated by predicate execution, aggregating the result on a single scalar value.
Q18 is dominated by a high-cardinality grouped aggregation.
Q03 and Q09 are very join-intensive, as most of the TPC-H queries.

\section{Analysis of Java Stream Performance}\label{sec:eval}
In this section we describe our performance analysis on JEDI.
First, we present the research questions addressed by our work (Sec.~\ref{sec:eval:rq}).
Then, we present our evaluation settings (Sec.~\ref{sec:eval:setting}).
Finally, we answer these questions, explaining the conclusions we derive from our performance analysis (Sec.~\ref{sec:eval:options}--\ref{sec:eval:complexity_metrics}).

\subsection{Research Questions}\label{sec:eval:rq}
Our performance analysis of the Java Stream API aims at answering the following research questions:

\begin{enumerate}
    \item[RQ1] Do the sequential options listed in Sec.~\ref{sec:codegen_options} (O1, O2) improve performance?
    \item[RQ2] Is there a parallelization strategy that always wins? Otherwise, which aspects should developers consider to select one of the proposed parallelization strategies?
    \item[RQ3] What is the performance of the Stream API w.r.t.~imperative code?
    \item[RQ4] Does the complexity of stream-based code differ from imperative code?
\end{enumerate}

\subsection{Evaluation Settings}\label{sec:eval:setting}
We perform our evaluation on two machines, henceforth called \emph{Intel} and \emph{ARM}.
Intel has a 16-core Intel Xeon~Gold~6326 CPU~@2.9GHz with 256GB~RAM~@3.2GHz, running the OS Ubuntu~22.04~LTS (kernel~5.15.0-25-generic).
Hyper-threading and turbo boost are disabled to ensure stable performance measurements.
ARM has a 80-core ARM~Neoverse-N1 CPU~@3.0GHz with 128GB~RAM~@3.2GHz, running Ubuntu~22.04 LTS (kernel~5.15.0-118-generic). 
We use two JVMs, henceforth called \emph{JDK} (Oracle JDK~24~\cite{oracle-jdk}, build~24.0.1+9-30) and \emph{GraalVM} (GraalVM~24\cite{graalvm,oracle-graalvm}, build~24.0.1+9.1).

Our generated benchmarks rely on the Java Microbenchmark Harness (JMH)~\cite{jmh}, a state-of-the-art Java harness library that eases the creation and execution of benchmarks on the JVM.
For all experiments, we execute 5~warmup and 5~measured iterations, each iteration lasting for 10~seconds.
In this way, we ensure evaluating performance in steady state, after the ergonomics phase~\cite{oracle-ergonomics} ends, i.e., the compiler, the garbage collector, and others JVM components that perform behavior-based tuning of the running application have completed.
We make sure to avoid any pitfall of JMH usage~\cite{jmh-do-dons-costa}.

\subsection{Stream Options (RQ1)}\label{sec:eval:options}

\begin{figure*}[t]
    \setlength{\tabcolsep}{1.6pt}
    \centering
    \begin{minipage}{.48\textwidth}
        \centering
        \begin{tabular}{lr|r|r|r||r|r|r|r}
\toprule
& \multicolumn{4}{c||}{\textbf{JDK, Intel}} & \multicolumn{4}{c}{\textbf{JDK, ARM}} \\ 
\midrule
& \multicolumn{1}{c|}{\textbf{B [ms]}} & \multicolumn{1}{c|}{\textbf{O1}} & \multicolumn{1}{c|}{\textbf{O2}} & \multicolumn{1}{c||}{\textbf{Both}}& \multicolumn{1}{c|}{\textbf{B [ms]}} & \multicolumn{1}{c|}{\textbf{O1}} & \multicolumn{1}{c|}{\textbf{O2}} & \multicolumn{1}{c}{\textbf{Both}} \\ 
\midrule
\textbf{Q01} & 363$\pm$4 & \textbf{=B} & \textbf{=B} & \textbf{=B} & 616$\pm$3 & \textbf{=B} & \textbf{=B} & \textbf{=B} \\ 
\textbf{Q02} & 182$\pm$2 & \textbf{1.02}x & \textbf{2.28}x & \textbf{2.38}x & 289$\pm$8 & \textbf{0.99}x & \textbf{1.98}x & \textbf{2.05}x \\ 
\textbf{Q03} & 566$\pm$5 & \textbf{=B} & \textbf{1.78}x & \textbf{=O2} & 908$\pm$13 & \textbf{=B} & \textbf{1.42}x & \textbf{=O2} \\ 
\textbf{Q04} & 405$\pm$4 & \textbf{1.04}x & \textbf{=B} & \textbf{=O1} & 607$\pm$5 & \textbf{0.98}x & \textbf{=B} & \textbf{=O1} \\ 
\textbf{Q05} & 818$\pm$23 & \textbf{1.03}x & \textbf{2.39}x & \textbf{2.62}x & 1162$\pm$5 & \textbf{1.00}x & \textbf{1.68}x & \textbf{1.74}x \\ 
\textbf{Q06} & 186$\pm$0 & \textbf{1.55}x & \textbf{=B} & \textbf{=O1} & 347$\pm$8 & \textbf{1.74}x & \textbf{=B} & \textbf{=O1} \\ 
\textbf{Q07} & 568$\pm$6 & \textbf{1.11}x & \textbf{1.52}x & \textbf{1.93}x & 925$\pm$20 & \textbf{1.17}x & \textbf{1.31}x & \textbf{1.59}x \\ 
\textbf{Q08} & 879$\pm$17 & \textbf{0.96}x & \textbf{2.29}x & \textbf{2.39}x & 1184$\pm$6 & \textbf{1.01}x & \textbf{1.67}x & \textbf{1.68}x \\ 
\textbf{Q09} & 1562$\pm$41 & \textbf{=B} & \textbf{1.94}x & \textbf{=O2} & 1992$\pm$33 & \textbf{=B} & \textbf{1.61}x & \textbf{=O2} \\ 
\textbf{Q10} & 448$\pm$9 & \textbf{1.10}x & \textbf{1.21}x & \textbf{1.26}x & 670$\pm$16 & \textbf{1.13}x & \textbf{1.07}x & \textbf{1.20}x \\ 
\textbf{Q11} & 111$\pm$0 & \textbf{=B} & \textbf{2.39}x & \textbf{=O2} & 197$\pm$2 & \textbf{=B} & \textbf{2.11}x & \textbf{=O2} \\ 
\textbf{Q12} & 399$\pm$3 & \textbf{1.16}x & \textbf{1.14}x & \textbf{1.35}x & 668$\pm$61 & \textbf{1.19}x & \textbf{1.10}x & \textbf{1.36}x \\ 
\textbf{Q13} & 450$\pm$40 & \textbf{=B} & \textbf{1.05}x & \textbf{=O2} & 696$\pm$5 & \textbf{=B} & \textbf{1.02}x & \textbf{=O2} \\ 
\textbf{Q14} & 201$\pm$1 & \textbf{1.03}x & \textbf{1.08}x & \textbf{1.13}x & 414$\pm$10 & \textbf{1.02}x & \textbf{1.05}x & \textbf{1.07}x \\ 
\textbf{Q15} & 383$\pm$3 & \textbf{1.14}x & \textbf{1.00}x & \textbf{1.16}x & 876$\pm$22 & \textbf{1.17}x & \textbf{0.99}x & \textbf{1.06}x \\ 
\textbf{Q16} & 130$\pm$0 & \textbf{1.02}x & \textbf{1.25}x & \textbf{1.27}x & 208$\pm$1 & \textbf{1.02}x & \textbf{1.31}x & \textbf{1.34}x \\ 
\textbf{Q17} & 1320$\pm$23 & \textbf{1.00}x & \textbf{1.56}x & \textbf{1.56}x & 1713$\pm$26 & \textbf{0.98}x & \textbf{1.29}x & \textbf{1.32}x \\ 
\textbf{Q18} & 1005$\pm$61 & \textbf{=B} & \textbf{1.53}x & \textbf{=O2} & 1533$\pm$27 & \textbf{=B} & \textbf{1.28}x & \textbf{=O2} \\ 
\textbf{Q19} & 338$\pm$2 & \textbf{1.14}x & \textbf{1.14}x & \textbf{1.23}x & 555$\pm$10 & \textbf{1.12}x & \textbf{1.16}x & \textbf{1.17}x \\ 
\textbf{Q20} & 708$\pm$10 & \textbf{1.10}x & \textbf{1.15}x & \textbf{1.30}x & 1026$\pm$5 & \textbf{1.08}x & \textbf{1.06}x & \textbf{1.21}x \\ 
\textbf{Q21} & 3021$\pm$160 & \textbf{=B} & \textbf{1.34}x & \textbf{=O2} & 4562$\pm$86 & \textbf{=B} & \textbf{1.16}x & \textbf{=O2} \\ 
\textbf{Q22} & 173$\pm$3 & \textbf{1.03}x & \textbf{1.00}x & \textbf{1.00}x & 283$\pm$4 & \textbf{1.02}x & \textbf{1.00}x & \textbf{1.04}x \\ 
 \hline 
 \multicolumn{2}{l|}{\textbf{Geo Mean}}  & \textbf{1.06x} & \textbf{1.38x} & \textbf{1.48x} &  & \textbf{1.06x} & \textbf{1.25x} & \textbf{1.34x} \\ 
\bottomrule
\end{tabular}
    \end{minipage}%
    \hfill
    \begin{minipage}{.48\textwidth}
        \centering
        \begin{tabular}{lr|r|r|r||r|r|r|r}
\toprule
& \multicolumn{4}{c||}{\textbf{GraalVM, Intel}} & \multicolumn{4}{c}{\textbf{GraalVM, ARM}} \\ 
\midrule
& \multicolumn{1}{c|}{\textbf{B [ms]}} & \multicolumn{1}{c|}{\textbf{O1}} & \multicolumn{1}{c|}{\textbf{O2}} & \multicolumn{1}{c||}{\textbf{Both}}& \multicolumn{1}{c|}{\textbf{B [ms]}} & \multicolumn{1}{c|}{\textbf{O1}} & \multicolumn{1}{c|}{\textbf{O2}} & \multicolumn{1}{c}{\textbf{Both}} \\ 
\midrule
\textbf{Q01} & 333$\pm$4 & \textbf{=B} & \textbf{=B} & \textbf{=B} & 558$\pm$11 & \textbf{=B} & \textbf{=B} & \textbf{=B} \\ 
\textbf{Q02} & 128$\pm$1 & \textbf{0.97}x & \textbf{1.85}x & \textbf{1.89}x & 228$\pm$15 & \textbf{1.01}x & \textbf{1.68}x & \textbf{1.72}x \\ 
\textbf{Q03} & 452$\pm$5 & \textbf{=B} & \textbf{1.46}x & \textbf{=O2} & 807$\pm$5 & \textbf{=B} & \textbf{1.26}x & \textbf{=O2} \\ 
\textbf{Q04} & 348$\pm$4 & \textbf{1.05}x & \textbf{=B} & \textbf{=O1} & 668$\pm$5 & \textbf{1.13}x & \textbf{=B} & \textbf{=O1} \\ 
\textbf{Q05} & 519$\pm$19 & \textbf{1.04}x & \textbf{2.06}x & \textbf{2.10}x & 862$\pm$7 & \textbf{1.07}x & \textbf{1.52}x & \textbf{1.76}x \\ 
\textbf{Q06} & 158$\pm$0 & \textbf{1.15}x & \textbf{=B} & \textbf{=O1} & 336$\pm$2 & \textbf{1.71}x & \textbf{=B} & \textbf{=O1} \\ 
\textbf{Q07} & 436$\pm$9 & \textbf{1.08}x & \textbf{1.62}x & \textbf{1.86}x & 787$\pm$11 & \textbf{1.08}x & \textbf{1.43}x & \textbf{1.55}x \\ 
\textbf{Q08} & 594$\pm$13 & \textbf{1.05}x & \textbf{1.89}x & \textbf{2.02}x & 872$\pm$16 & \textbf{1.09}x & \textbf{1.61}x & \textbf{1.70}x \\ 
\textbf{Q09} & 920$\pm$9 & \textbf{=B} & \textbf{2.16}x & \textbf{=O2} & 1523$\pm$39 & \textbf{=B} & \textbf{1.69}x & \textbf{=O2} \\ 
\textbf{Q10} & 397$\pm$10 & \textbf{1.01}x & \textbf{1.22}x & \textbf{1.31}x & 623$\pm$16 & \textbf{1.11}x & \textbf{1.12}x & \textbf{1.18}x \\ 
\textbf{Q11} & 82$\pm$2 & \textbf{=B} & \textbf{2.15}x & \textbf{=O2} & 163$\pm$2 & \textbf{=B} & \textbf{1.85}x & \textbf{=O2} \\ 
\textbf{Q12} & 274$\pm$3 & \textbf{1.01}x & \textbf{0.98}x & \textbf{1.02}x & 528$\pm$57 & \textbf{1.00}x & \textbf{0.96}x & \textbf{1.02}x \\ 
\textbf{Q13} & 296$\pm$5 & \textbf{=B} & \textbf{1.02}x & \textbf{=O2} & 567$\pm$11 & \textbf{=B} & \textbf{1.01}x & \textbf{=O2} \\ 
\textbf{Q14} & 167$\pm$0 & \textbf{1.04}x & \textbf{1.03}x & \textbf{1.08}x & 338$\pm$8 & \textbf{1.20}x & \textbf{1.00}x & \textbf{1.26}x \\ 
\textbf{Q15} & 334$\pm$3 & \textbf{1.14}x & \textbf{0.99}x & \textbf{1.15}x & 723$\pm$49 & \textbf{1.38}x & \textbf{1.02}x & \textbf{1.26}x \\ 
\textbf{Q16} & 90$\pm$1 & \textbf{1.01}x & \textbf{1.03}x & \textbf{1.05}x & 166$\pm$2 & \textbf{1.03}x & \textbf{1.11}x & \textbf{1.13}x \\ 
\textbf{Q17} & 934$\pm$14 & \textbf{0.98}x & \textbf{1.63}x & \textbf{1.59}x & 1484$\pm$85 & \textbf{0.99}x & \textbf{1.26}x & \textbf{1.50}x \\ 
\textbf{Q18} & 720$\pm$28 & \textbf{=B} & \textbf{1.31}x & \textbf{=O2} & 1262$\pm$57 & \textbf{=B} & \textbf{1.09}x & \textbf{=O2} \\ 
\textbf{Q19} & 257$\pm$1 & \textbf{1.01}x & \textbf{1.07}x & \textbf{1.03}x & 525$\pm$5 & \textbf{1.05}x & \textbf{1.02}x & \textbf{1.07}x \\ 
\textbf{Q20} & 479$\pm$14 & \textbf{1.05}x & \textbf{1.23}x & \textbf{1.38}x & 859$\pm$102 & \textbf{1.09}x & \textbf{1.16}x & \textbf{1.26}x \\ 
\textbf{Q21} & 2537$\pm$34 & \textbf{=B} & \textbf{1.30}x & \textbf{=O2} & 4161$\pm$37 & \textbf{=B} & \textbf{1.15}x & \textbf{=O2} \\ 
\textbf{Q22} & 125$\pm$4 & \textbf{1.01}x & \textbf{1.03}x & \textbf{1.00}x & 251$\pm$4 & \textbf{1.09}x & \textbf{1.01}x & \textbf{1.09}x \\ 
 \hline 
 \multicolumn{2}{l|}{\textbf{Geo Mean}}  & \textbf{1.03x} & \textbf{1.31x} & \textbf{1.36x} &  & \textbf{1.08x} & \textbf{1.20x} & \textbf{1.31x} \\ 
\bottomrule
\end{tabular}
    \end{minipage}

    \captionsetup{type=table}
    \caption{Execution time [ms] with margin of error (half of the confidence interval) for the baseline (B) and speedup factor of O1, O2 and both, for JEDI queries on JDK (left) and GraalVM (right). 
    "=B" means the option is not applicable (i.e., same code as baseline). "=O1" ("=O2"), means only O1 (O2) is applicable. 
    The last row reports the geometric mean of the speedups. SF$=$1.}\vspace{-3mm}
    \label{tab:options}
\end{figure*}

Here, we answer to RQ1, focusing on the first two options described in Sec.~\ref{sec:codegen_options}: filter fusion (O1) and join implementation with \code{mapMulti} (O2).
We run the experiments using a dataset of~1GB (\mbox{SF$=$1}).
Table~\ref{tab:options} shows the results for JDK (left) and GraalVM (right); we report the results on both Intel and ARM.
If a given option does not affect the generated code, we indicate "=B" (baseline), "=O1", or "=O2".

As the tables show, both options have a positive performance impact. 
As joins dominate the execution time for most queries~\cite{chokepoints1}, O2 is usually more effective than O1.
On JDK, O2 results in speedup factors from 0.99x (Q15, on ARM) to 2.39x (Q5 and Q11, on Intel), with a geometric mean of 1.38x on Intel and 1.25x on ARM. 
On the other hand, O1 results in speedup factors from 0.96x (Q8, on Intel) to 1.74x (Q06, on ARM), with a geometric mean of 1.06x (both machines). 
As O1 and O2 are orthogonal, applying them together (column ``Both'' in the tables, ``O1+O2'' in the following text) 
yields speedup factors from 1.00x (Q22, on Intel) to 2.62x (Q5, on Intel), with a geometric mean of 1.48x on Intel and 1.34x on ARM.

The experiments show similar trends on GraalVM (Table~\ref{tab:options}, right).
The most noticeable difference concerns O1 on Q06, where the speedup on GraalVM (1.15x) is lower than on JDK (1.55x). 
The reason is that GraalVM often inlines more methods than JDK, resulting in larger compilation units that are amenable to more optimizations. 
If the JIT compiler inlines some calls to \code{filter}, the resulting machine code becomes similar to the code obtained with O1.

For Q2 (Intel) and Q17 (both machines), there are a minor slowdowns when using O1 on GraalVM, there is also a minor slowdown for Q12 when using O2 (ARM). 
As these slowdowns are less than 5\% of the execution time, we consider them insignificant---they may be attributed to different heuristic decisions by the JIT~compiler.

Regarding O1, we also note that TPC-H makes use of a conjunction of a large number of predicates only in Q06 (5~predicates), and on this query, we can indeed see high benefits using O1 (1.55x on JDK, 1.15x on GraalVM with Intel and 1.74x on JDK, 1.71x on GraalVM with ARM).
To further confirm the benefit of O1, we evaluate the execution time on a query that we create as a microbenchmark to stress O1.
Figure~\ref{fig:microquery-o1} shows this query, which uses 7~predicates (all of them do not filter any row) and a final \code{COUNT(*)} aggregation that can be executed very efficiently.
This query stresses predicate execution and hence is a good candidate for evaluating O1. 
On Intel and JDK, this query has an execution time of 379ms without O1, and 107ms with O1, yielding a speedup factor of 3.54x. 
On Intel and GraalVM, the same query has an execution time of 427ms without O1, and 128ms with O1, yielding a speedup factor of 3.33x.

\begin{figure}[h]
    \begin{verbatim}
SELECT COUNT(*) AS counter FROM lineitem
WHERE l_orderkey >= 0 AND l_linenumber >= 0
AND l_quantity >= 0 AND l_extendedprice >= 0
AND l_suppkey >= 0 AND l_partkey >= 0 AND l_tax >= 0
    \end{verbatim}
    \vspace{-5mm}
    \caption{Query with 7 conjuncted predicates.}
    \label{fig:microquery-o1}
\end{figure}

\textbf{Answer to RQ1:}
A performance analysis of O1 and O2 on JEDI shows that both options improve performance. 
Thus, we suggest developers to apply both O1 and O2 whenever possible.

\subsection{Parallel Streams (RQ2)}\label{sec:eval:stream_parallel}

\begin{table*}[!t]
    \centering
    
    \begin{minipage}{.49\textwidth}
        \centering
        \makebox[\textwidth][c]{\textbf{JDK, Intel}}
        \vspace{3pt}
        \begin{tabular}{L{10pt}R{50pt}|R{30pt}|R{30pt}|R{30pt}|R{30pt}}
\toprule
& \textbf{Baseline} & \multicolumn{1}{c|}{\textbf{P}} & \multicolumn{1}{c|}{\textbf{PU}} & \multicolumn{1}{c|}{\textbf{CG}} & \multicolumn{1}{c}{\textbf{CGCC}} \\ 
\midrule
\textbf{Q01} & 3854$\pm$18 & \textbf{8.15x} & \textbf{8.13x} & \textbf{0.42x} & \textbf{1.42x} \\ 
\textbf{Q02} & 921$\pm$165 & \textbf{2.76x} & \textbf{2.79x} & \textbf{10.28x} & \textbf{10.48x} \\ 
\textbf{Q03} & 5171$\pm$83 & \textbf{7.39x} & \textbf{7.40x} & \textbf{9.91x} & \textbf{9.71x} \\ 
\textbf{Q04} & 4130$\pm$288 & \textbf{2.23x} & \textbf{2.24x} & \textbf{2.21x} & \textbf{2.22x} \\ 
\textbf{Q05} & 3956$\pm$60 & \textbf{9.78x} & \textbf{9.84x} & \textbf{11.68x} & \textbf{11.67x} \\ 
\textbf{Q06} & 1257$\pm$4 & \textbf{9.70x} & \textbf{9.97x} & \textbf{9.60x} & \textbf{10.28x} \\ 
\textbf{Q07} & 3625$\pm$89 & \textbf{7.22x} & \textbf{7.22x} & \textbf{9.33x} & \textbf{9.75x} \\ 
\textbf{Q08} & 3856$\pm$47 & \textbf{9.94x} & \textbf{9.56x} & \textbf{11.72x} & \textbf{12.47x} \\ 
\textbf{Q09} & 13950$\pm$571 & \textbf{8.09x} & \textbf{8.15x} & \textbf{10.85x} & \textbf{11.30x} \\ 
\textbf{Q10} & 4541$\pm$97 & \textbf{5.28x} & \textbf{5.37x} & \textbf{10.52x} & \textbf{9.86x} \\ 
\textbf{Q11} & 489$\pm$1 & \textbf{4.63x} & \textbf{4.62x} & \textbf{9.20x} & \textbf{9.24x} \\ 
\textbf{Q12} & 3264$\pm$55 & \textbf{9.86x} & \textbf{9.88x} & \textbf{10.99x} & \textbf{10.73x} \\ 
\textbf{Q13} & 6044$\pm$84 & \textbf{2.32x} & \textbf{2.33x} & \textbf{8.25x} & \textbf{7.56x} \\ 
\textbf{Q14} & 1985$\pm$31 & \textbf{6.61x} & \textbf{6.42x} & \textbf{9.89x} & \textbf{11.99x} \\ 
\textbf{Q16} & 1192$\pm$45 & \textbf{2.25x} & \textbf{2.27x} & \textbf{6.75x} & \textbf{6.72x} \\ 
\textbf{Q17} & 12328$\pm$765 & \textbf{2.90x} & \textbf{2.98x} & \textbf{8.49x} & \textbf{8.29x} \\ 
\textbf{Q18} & 7156$\pm$162 & \textbf{2.89x} & \textbf{2.86x} & \textbf{4.70x} & \textbf{4.58x} \\ 
\textbf{Q19} & 3221$\pm$2 & \textbf{4.28x} & \textbf{4.23x} & \textbf{11.05x} & \textbf{10.72x} \\ 
\textbf{Q20} & 7290$\pm$204 & \textbf{2.42x} & \textbf{2.45x} & \textbf{7.80x} & \textbf{7.67x} \\ 
\textbf{Q21} & 29965$\pm$1586 & \textbf{3.80x} & \textbf{3.80x} & \textbf{5.73x} & \textbf{5.34x} \\ 
\textbf{Q22} & 3236$\pm$69 & \textbf{2.52x} & \textbf{2.50x} & \textbf{7.83x} & \textbf{8.34x} \\ 
 \hline 
 \multicolumn{2}{l|}{\textbf{Geo Mean}} &\textbf{4.72x} & \textbf{4.72x} & \textbf{7.24x} & \textbf{7.72x} \\ 
\bottomrule
\end{tabular}
    \end{minipage}%
    \hfill
    \begin{minipage}{.49\textwidth}
        \centering
        \makebox[\textwidth][c]{\textbf{GraalVM, Intel}}
        \vspace{3pt}
        \begin{tabular}{L{10pt}R{50pt}|R{30pt}|R{30pt}|R{30pt}|R{30pt}}
\toprule
& \textbf{Baseline} & \multicolumn{1}{c|}{\textbf{P}} & \multicolumn{1}{c|}{\textbf{PU}} & \multicolumn{1}{c|}{\textbf{CG}} & \multicolumn{1}{c}{\textbf{CGCC}} \\ 
\midrule
\textbf{Q01} & 3578$\pm$19 & \textbf{8.47x} & \textbf{8.45x} & \textbf{0.43x} & \textbf{1.18x} \\ 
\textbf{Q02} & 785$\pm$30 & \textbf{2.69x} & \textbf{2.81x} & \textbf{8.74x} & \textbf{4.21x} \\ 
\textbf{Q03} & 5328$\pm$143 & \textbf{8.01x} & \textbf{7.99x} & \textbf{9.90x} & \textbf{9.78x} \\ 
\textbf{Q04} & 4056$\pm$270 & \textbf{2.19x} & \textbf{2.03x} & \textbf{2.10x} & \textbf{2.10x} \\ 
\textbf{Q05} & 3076$\pm$37 & \textbf{9.53x} & \textbf{9.57x} & \textbf{11.34x} & \textbf{11.74x} \\ 
\textbf{Q06} & 1466$\pm$22 & \textbf{10.48x} & \textbf{10.62x} & \textbf{10.45x} & \textbf{10.45x} \\ 
\textbf{Q07} & 4448$\pm$185 & \textbf{9.87x} & \textbf{9.29x} & \textbf{12.51x} & \textbf{12.53x} \\ 
\textbf{Q08} & 3361$\pm$15 & \textbf{9.85x} & \textbf{10.38x} & \textbf{12.94x} & \textbf{12.50x} \\ 
\textbf{Q09} & 9347$\pm$164 & \textbf{7.74x} & \textbf{7.60x} & \textbf{11.15x} & \textbf{11.05x} \\ 
\textbf{Q10} & 3878$\pm$112 & \textbf{5.53x} & \textbf{5.56x} & \textbf{9.73x} & \textbf{9.09x} \\ 
\textbf{Q11} & 455$\pm$14 & \textbf{5.60x} & \textbf{5.68x} & \textbf{10.13x} & \textbf{9.45x} \\ 
\textbf{Q12} & 3011$\pm$67 & \textbf{9.57x} & \textbf{9.45x} & \textbf{9.97x} & \textbf{10.13x} \\ 
\textbf{Q13} & 4323$\pm$170 & \textbf{1.93x} & \textbf{1.97x} & \textbf{6.92x} & \textbf{7.68x} \\ 
\textbf{Q14} & 1678$\pm$2 & \textbf{6.23x} & \textbf{6.34x} & \textbf{9.27x} & \textbf{8.83x} \\ 
\textbf{Q16} & 1067$\pm$55 & \textbf{2.44x} & \textbf{2.42x} & \textbf{6.65x} & \textbf{6.88x} \\ 
\textbf{Q17} & 8942$\pm$53 & \textbf{2.63x} & \textbf{2.68x} & \textbf{7.72x} & \textbf{7.15x} \\ 
\textbf{Q18} & 5307$\pm$523 & \textbf{2.64x} & \textbf{2.49x} & \textbf{4.14x} & \textbf{3.93x} \\ 
\textbf{Q19} & 2911$\pm$36 & \textbf{4.41x} & \textbf{4.28x} & \textbf{9.59x} & \textbf{9.68x} \\ 
\textbf{Q20} & 4861$\pm$552 & \textbf{2.14x} & \textbf{2.13x} & \textbf{6.95x} & \textbf{6.66x} \\ 
\textbf{Q21} & 27519$\pm$804 & \textbf{4.05x} & \textbf{4.06x} & \textbf{5.81x} & \textbf{5.39x} \\ 
\textbf{Q22} & 2184$\pm$63 & \textbf{2.08x} & \textbf{2.06x} & \textbf{6.76x} & \textbf{6.93x} \\ 
 \hline 
 \multicolumn{2}{l|}{\textbf{Geo Mean}} &\textbf{4.72x} & \textbf{4.70x} & \textbf{7.02x} & \textbf{7.02x} \\ 
\bottomrule
\end{tabular}
    \end{minipage}
    \vspace{8pt}

    \vspace{2pt}
    \begin{minipage}{.49\textwidth}
        \centering
        \makebox[\textwidth][c]{\textbf{JDK, ARM}}
        \vspace{2pt}
        \begin{tabular}{L{10pt}R{50pt}|R{30pt}|R{30pt}|R{30pt}|R{30pt}}
\toprule
& \textbf{Baseline} & \multicolumn{1}{c|}{\textbf{P}} & \multicolumn{1}{c|}{\textbf{PU}} & \multicolumn{1}{c|}{\textbf{CG}} & \multicolumn{1}{c}{\textbf{CGCC}} \\ 
\midrule
\textbf{Q01} & 6030$\pm$87 & \textbf{13.79x} & \textbf{13.68x} & \textbf{0.44x} & \textbf{1.62x} \\ 
\textbf{Q02} & 1869$\pm$71 & \textbf{3.90x} & \textbf{3.92x} & \textbf{18.24x} & \textbf{18.12x} \\ 
\textbf{Q03} & 8755$\pm$296 & \textbf{9.58x} & \textbf{9.71x} & \textbf{17.27x} & \textbf{17.61x} \\ 
\textbf{Q04} & 7826$\pm$267 & \textbf{3.44x} & \textbf{3.41x} & \textbf{3.29x} & \textbf{3.40x} \\ 
\textbf{Q05} & 7460$\pm$208 & \textbf{19.00x} & \textbf{19.29x} & \textbf{29.98x} & \textbf{29.77x} \\ 
\textbf{Q06} & 1810$\pm$11 & \textbf{6.35x} & \textbf{6.38x} & \textbf{6.35x} & \textbf{6.35x} \\ 
\textbf{Q07} & 6052$\pm$93 & \textbf{8.20x} & \textbf{8.33x} & \textbf{13.62x} & \textbf{13.45x} \\ 
\textbf{Q08} & 7215$\pm$300 & \textbf{19.91x} & \textbf{19.85x} & \textbf{27.42x} & \textbf{26.09x} \\ 
\textbf{Q09} & 20239$\pm$1472 & \textbf{12.65x} & \textbf{12.63x} & \textbf{25.34x} & \textbf{25.18x} \\ 
\textbf{Q10} & 6550$\pm$180 & \textbf{6.67x} & \textbf{6.70x} & \textbf{14.11x} & \textbf{13.14x} \\ 
\textbf{Q11} & 1081$\pm$33 & \textbf{7.54x} & \textbf{7.70x} & \textbf{20.87x} & \textbf{21.40x} \\ 
\textbf{Q12} & 5565$\pm$34 & \textbf{13.68x} & \textbf{14.00x} & \textbf{12.57x} & \textbf{15.63x} \\ 
\textbf{Q13} & 8943$\pm$161 & \textbf{2.94x} & \textbf{2.98x} & \textbf{13.97x} & \textbf{23.38x} \\ 
\textbf{Q14} & 3863$\pm$119 & \textbf{7.85x} & \textbf{7.97x} & \textbf{12.46x} & \textbf{12.70x} \\ 
\textbf{Q16} & 1735$\pm$26 & \textbf{2.37x} & \textbf{2.38x} & \textbf{9.06x} & \textbf{8.64x} \\ 
\textbf{Q17} & 18697$\pm$549 & \textbf{3.33x} & \textbf{3.31x} & \textbf{16.84x} & \textbf{14.99x} \\ 
\textbf{Q18} & 13399$\pm$290 & \textbf{4.64x} & \textbf{4.65x} & \textbf{12.08x} & \textbf{11.32x} \\ 
\textbf{Q19} & 5374$\pm$109 & \textbf{5.44x} & \textbf{5.34x} & \textbf{15.43x} & \textbf{15.75x} \\ 
\textbf{Q20} & 10339$\pm$233 & \textbf{2.84x} & \textbf{2.83x} & \textbf{17.20x} & \textbf{16.53x} \\ 
\textbf{Q21} & 48741$\pm$482 & \textbf{4.93x} & \textbf{5.04x} & \textbf{12.45x} & \textbf{12.37x} \\ 
\textbf{Q22} & 4778$\pm$255 & \textbf{2.84x} & \textbf{2.85x} & \textbf{10.91x} & \textbf{11.53x} \\ 
 \hline 
 \multicolumn{2}{l|}{\textbf{Geo Mean}} &\textbf{6.27x} & \textbf{6.31x} & \textbf{11.85x} & \textbf{12.90x} \\ 
\bottomrule
\end{tabular}
    \end{minipage}%
    \hfill 
    \begin{minipage}{.49\textwidth}
        \centering
        \makebox[\textwidth][c]{\textbf{GraalVM, ARM}}
        \vspace{2pt}
        \hspace{-7pt}
        \begin{tabular}{L{10pt}R{50pt}|R{30pt}|R{30pt}|R{30pt}|R{30pt}}
\toprule
& \textbf{Baseline} & \multicolumn{1}{c|}{\textbf{P}} & \multicolumn{1}{c|}{\textbf{PU}} & \multicolumn{1}{c|}{\textbf{CG}} & \multicolumn{1}{c}{\textbf{CGCC}} \\ 
\midrule
\textbf{Q01} & 5139$\pm$122 & \textbf{14.34x} & \textbf{14.28x} & \textbf{0.36x} & \textbf{1.41x} \\ 
\textbf{Q02} & 1633$\pm$33 & \textbf{4.22x} & \textbf{4.40x} & \textbf{14.88x} & \textbf{9.90x} \\ 
\textbf{Q03} & 8538$\pm$336 & \textbf{10.15x} & \textbf{10.36x} & \textbf{19.67x} & \textbf{19.48x} \\ 
\textbf{Q04} & 6873$\pm$293 & \textbf{3.04x} & \textbf{3.04x} & \textbf{2.97x} & \textbf{2.92x} \\ 
\textbf{Q05} & 5687$\pm$105 & \textbf{16.52x} & \textbf{16.94x} & \textbf{25.67x} & \textbf{24.71x} \\ 
\textbf{Q06} & 2476$\pm$58 & \textbf{9.50x} & \textbf{8.72x} & \textbf{9.08x} & \textbf{9.10x} \\ 
\textbf{Q07} & 7505$\pm$724 & \textbf{11.36x} & \textbf{11.61x} & \textbf{19.39x} & \textbf{19.24x} \\ 
\textbf{Q08} & 6384$\pm$69 & \textbf{18.73x} & \textbf{18.93x} & \textbf{24.66x} & \textbf{25.50x} \\ 
\textbf{Q09} & 17256$\pm$1661 & \textbf{13.08x} & \textbf{12.86x} & \textbf{30.47x} & \textbf{30.75x} \\ 
\textbf{Q10} & 6436$\pm$512 & \textbf{7.11x} & \textbf{7.08x} & \textbf{13.62x} & \textbf{13.97x} \\ 
\textbf{Q11} & 973$\pm$4 & \textbf{8.87x} & \textbf{9.02x} & \textbf{19.82x} & \textbf{18.69x} \\ 
\textbf{Q12} & 6783$\pm$1310 & \textbf{17.08x} & \textbf{17.16x} & \textbf{14.66x} & \textbf{19.43x} \\ 
\textbf{Q13} & 8463$\pm$130 & \textbf{2.94x} & \textbf{2.94x} & \textbf{12.52x} & \textbf{21.58x} \\ 
\textbf{Q14} & 3221$\pm$247 & \textbf{7.27x} & \textbf{7.52x} & \textbf{11.00x} & \textbf{10.90x} \\ 
\textbf{Q16} & 1833$\pm$33 & \textbf{2.52x} & \textbf{2.51x} & \textbf{10.17x} & \textbf{11.15x} \\ 
\textbf{Q17} & 16647$\pm$761 & \textbf{2.98x} & \textbf{2.98x} & \textbf{16.94x} & \textbf{14.44x} \\ 
\textbf{Q18} & 10134$\pm$394 & \textbf{3.70x} & \textbf{3.87x} & \textbf{10.16x} & \textbf{8.72x} \\ 
\textbf{Q19} & 5241$\pm$155 & \textbf{5.85x} & \textbf{5.84x} & \textbf{15.79x} & \textbf{15.25x} \\ 
\textbf{Q20} & 9431$\pm$1024 & \textbf{2.99x} & \textbf{2.96x} & \textbf{18.47x} & \textbf{17.24x} \\ 
\textbf{Q21} & 44938$\pm$695 & \textbf{5.24x} & \textbf{5.11x} & \textbf{12.33x} & \textbf{11.43x} \\ 
\textbf{Q22} & 4730$\pm$139 & \textbf{2.34x} & \textbf{2.34x} & \textbf{11.26x} & \textbf{11.53x} \\ 
 \hline 
 \multicolumn{2}{l|}{\textbf{Geo Mean}} &\textbf{6.48x} & \textbf{6.51x} & \textbf{11.94x} & \textbf{12.72x} \\ 
\bottomrule
\end{tabular}
    \end{minipage}

    \caption{Execution time [ms] of the baseline with margin of error (half of the confidence interval) and speedup factor for each parallelization strategy, for JEDI queries on Intel (16 cores, top) and ARM (80 cores, bottom) on JDK (left) and GraalVM (right). Baseline: sequential stream (O1+O2); P: Parallel ordered stream; PU: parallel unordered stream; CG: PU with groupingByConcurrent; CGCC: CG with concurrent collector. SF$=$10.} 
    \label{tab:parallel}
\end{table*}

\begin{figure*}[t]
    \centering

    \begin{subfigure}[b]{\textwidth}
        \centering
        \includegraphics[width=.4\linewidth]{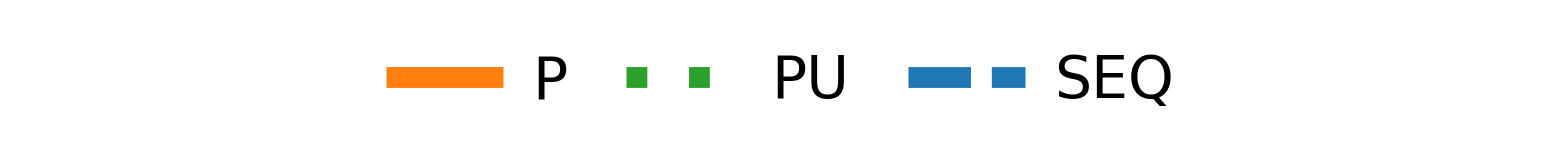}
    \end{subfigure}
    \begin{subfigure}[b]{0.47\textwidth}
        \centering
        \includegraphics[width=0.49\linewidth]{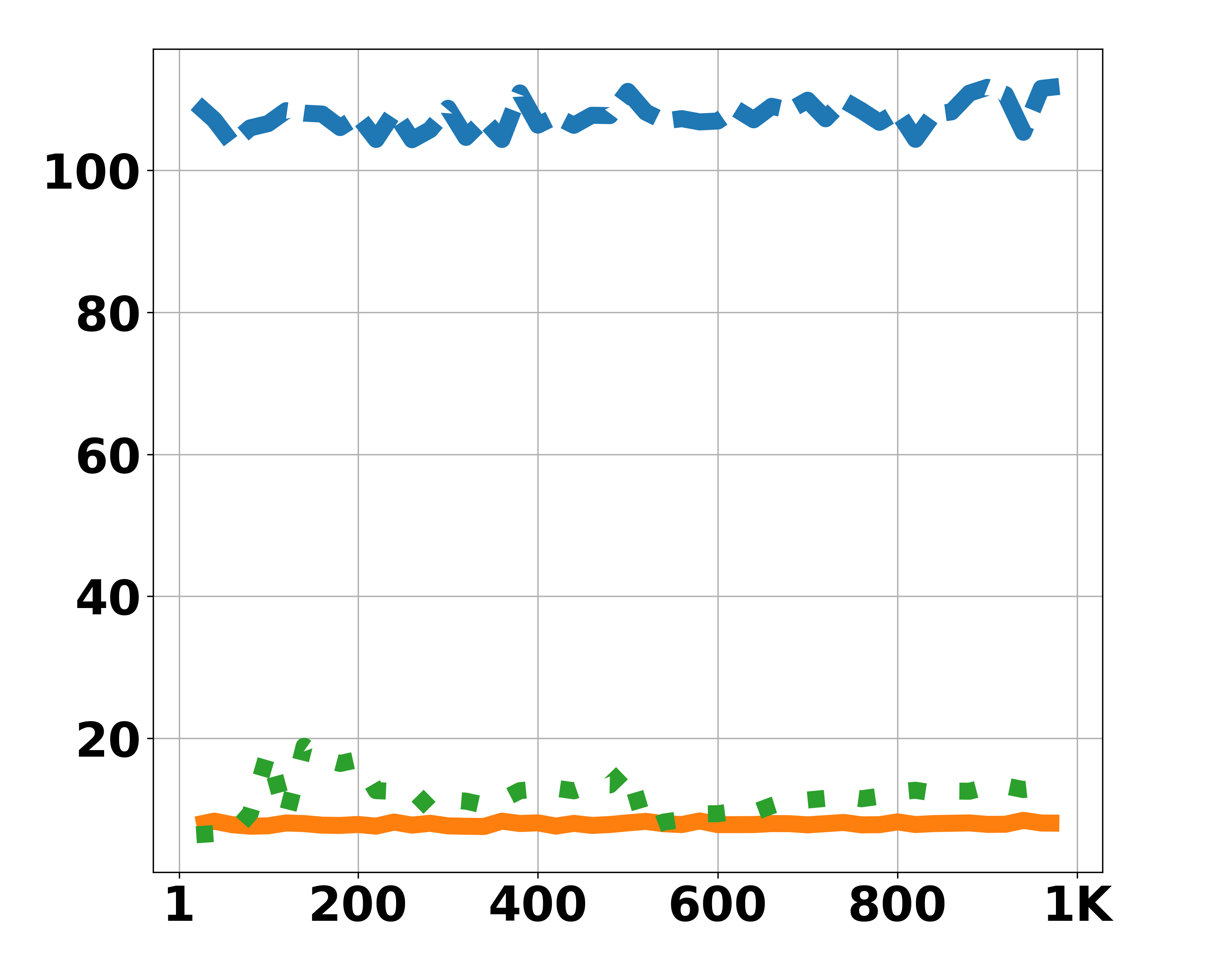}
        \includegraphics[width=0.49\linewidth]{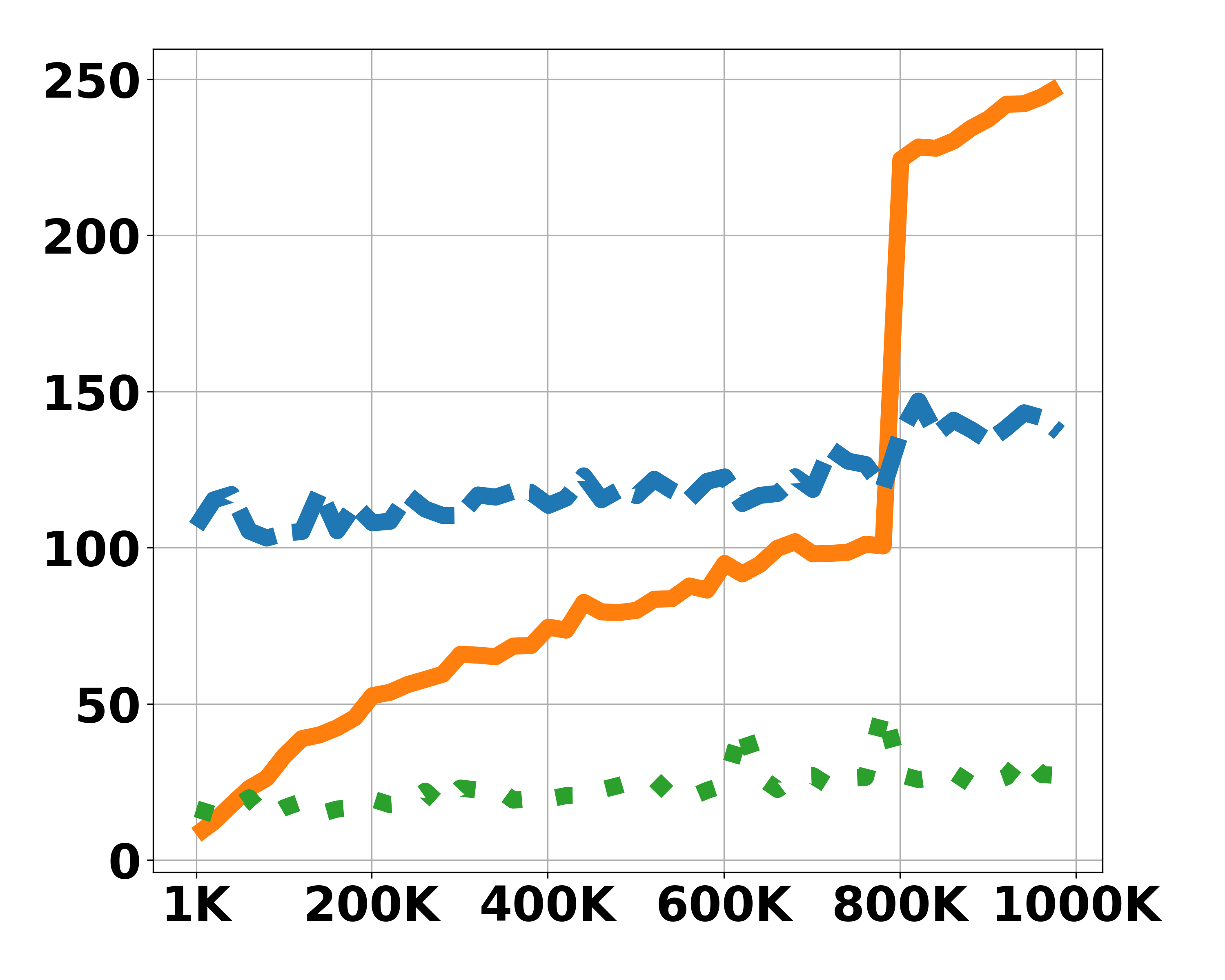}
        \caption{Intel}
    \end{subfigure}
    \begin{subfigure}[b]{0.02\textwidth}
        \centering
        \rule{0.3mm}{38mm}
    \end{subfigure}
    \begin{subfigure}[b]{0.47\textwidth}
        \centering
        \includegraphics[width=0.49\linewidth]{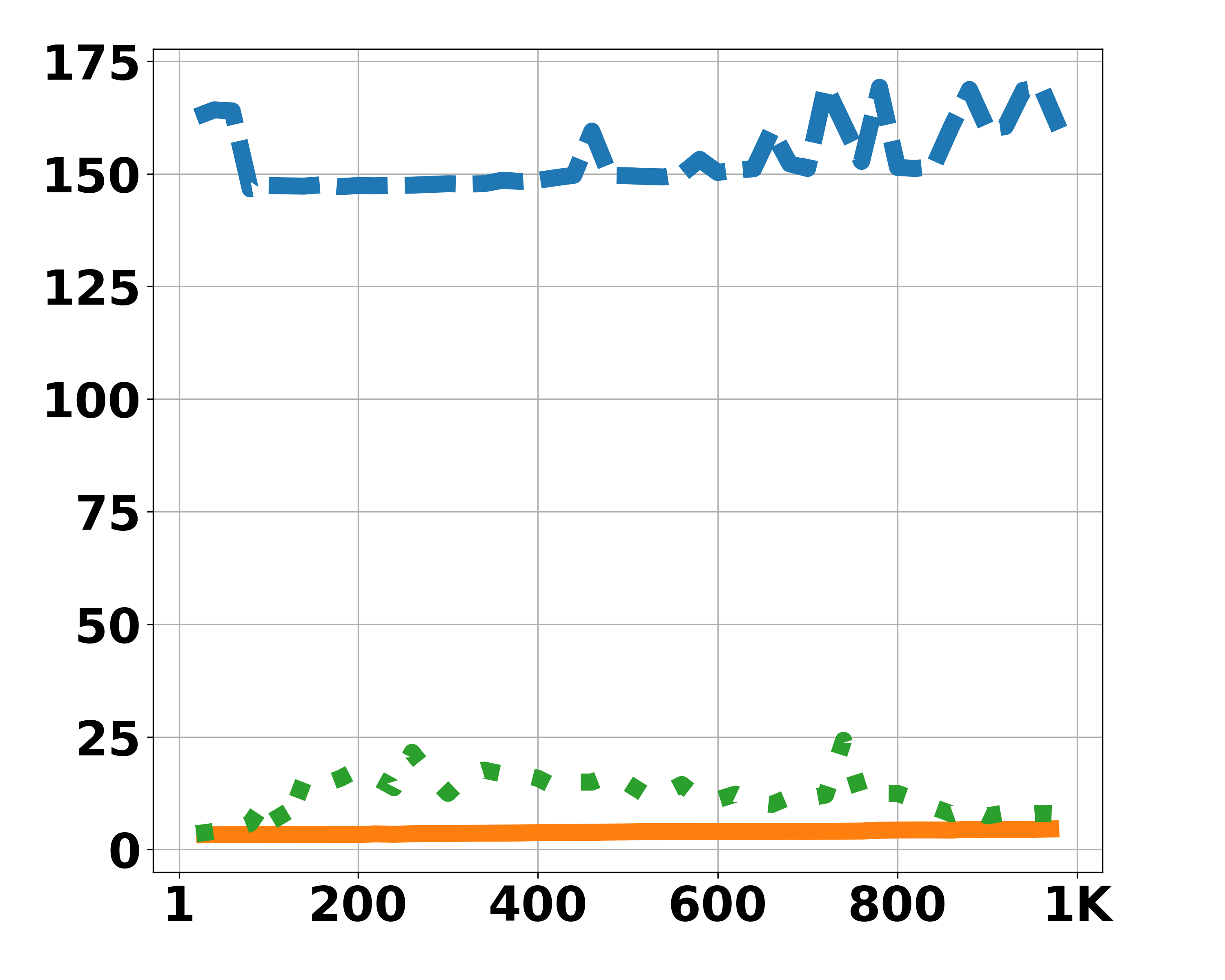}
        \includegraphics[width=0.49\linewidth]{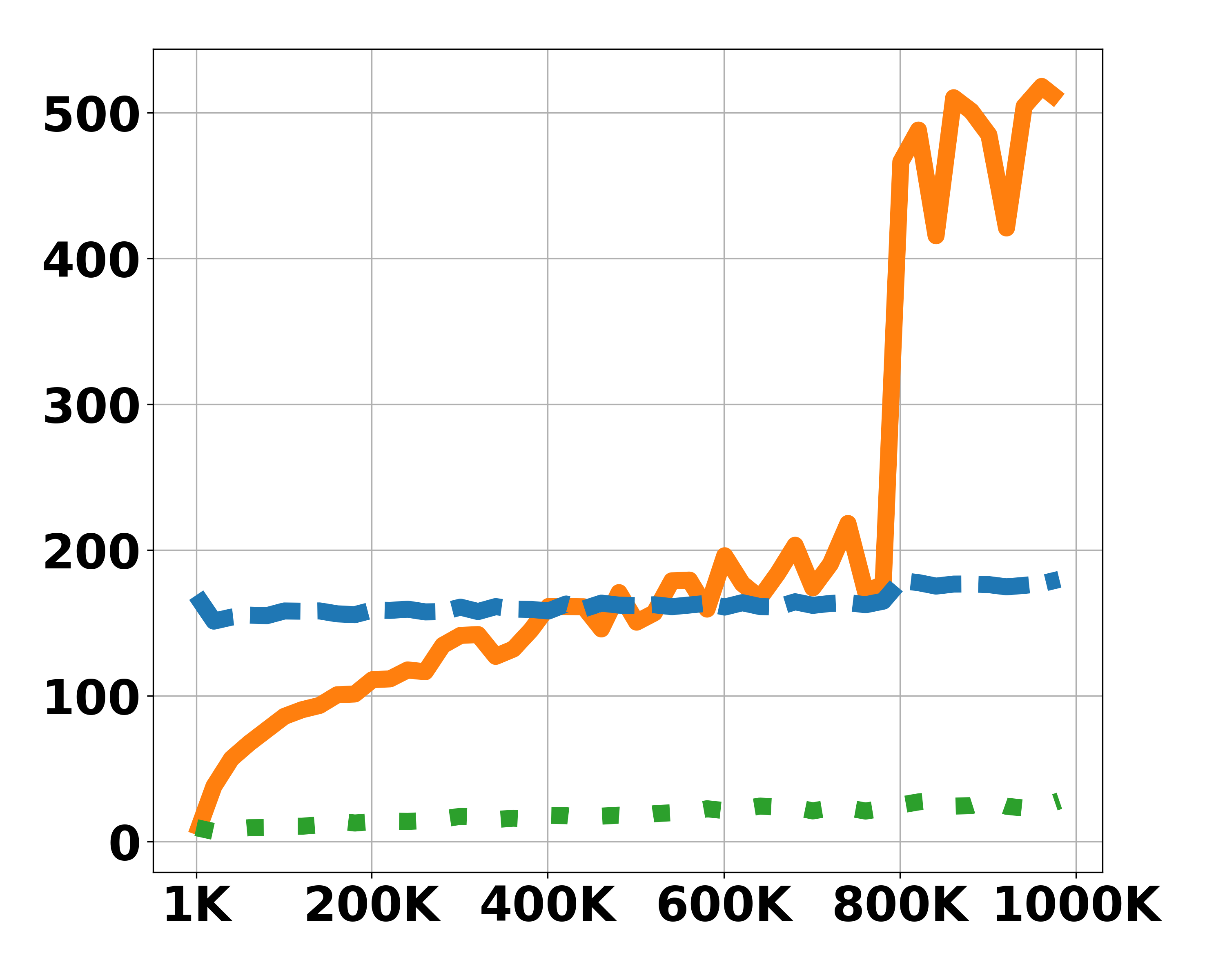}
        \caption{ARM}
    \end{subfigure}
    \caption{Execution time [ms] (y-axis) of the Distinct query varying the number of distinct elements (x-axis).}
    \label{fig:sio}
\end{figure*}

Here, we answer to RQ2, focusing on parallelization strategies for streams (O3 in Sec.~\ref{sec:codegen_options}).
Based on the lesson learnt from RQ1, we apply the parallelization strategies only on the best sequential implementation (O1+O2).
We run the experiments using a dataset of~10GB (\mbox{SF$=$10}).

Table~\ref{tab:parallel} shows the data collected on Intel (top) and on ARM (bottom) on both JDK (left) and GraalVM (right).
It shows the query execution time of the best sequential stream implementation (O1+O2, baseline) and the speedup factor for each parallelization strategy (O3) described in Sec.~\ref{sec:codegen_options}. 
The table excludes Q15, because our code generator uses doubles to represent decimal numbers.
Since Q15 uses the sum of decimals as a group-by key, the generated code may lead to a wrong result, because double addition is not associative, and parallelization may change the summation order.\footnote{We note that this problem could be solved by using Java's \code{BigDecimal} type, although this would introduce high overhead in a calculation that is unimportant for this work. We do not aim at implementing a real-world query engine, but at analyzing the performance of workloads stressing the Stream API.}

As the tables show, Q01 is an outlier.
First, it is the only query for which parallelization leads to a slowdown: 
the speedup of CG against the baseline is 0.42x on Intel and 0.44x on ARM using JDK.
Second, Q01 is the only query for which P and PU are much faster than CG and CGCC on both machines. 
Considering JDK, on Intel (Table~\ref{tab:parallel}, top-left) P and PU are faster than CG only on Q01 and Q06 (few ms), while on ARM (Table~\ref{tab:parallel}, bottom-left), PU is faster than CG only on Q01, Q04 (few ms), Q06 (few ms), and Q12.
For all other queries, P and PU are slower than CG and CGCC.

Surprisingly, P and PU perform similarly, even if PU has better parallelization opportunities since it can exploit unordered processing of the stream elements.
Unordered streams usually outperform ordered streams mostly in the presence of stateful intermediate operations (SIO), such as limit, skip, or distinct. 
TPC-H uses limit as a SIO, however, as commonly done in SQL queries, limit is used with sorting (i.e., queries in the form \code{SELECT X FROM T ORDER BY Y LIMIT N}). 
Such queries are converted into streams that use the sorted operation before limit; hence, the stream pipelines are ordered after the sorted operation, where limit is used, even if the previous segment of the pipeline was unordered.

To better highlight the performance of SIO with ordered and unordered streams, we present a microbenchmark focusing on the distinct operation, which is not used in TPC-H. 
For this experiment, we manually implement a stream pipeline that creates a list from the distinct elements in the data source, i.e., a SQL query like \code{SELECT DISTINCT(x) FROM orders}. 
Moreover, we artificially create the data source for this experiment such that we have full control over the number of distinct elements. The data source contains \SI{10000000} elements.
Figure~\ref{fig:sio} reports the results of this experiment. 
As expected~\cite{javadoc-distinct}, the distinct operation is usually faster on unordered streams.
This is expected, because the ordered version of distinct splits the computation in a fork-join parallel style which populates a \code{LinkedHashSet} for each task, joining pairs of such sets at each join phase, in contrast to the unordered version which populates a single \code{ConcurrentHashMap} shared by all threads.
As the figure shows, the fork-join implementation is particularly expensive when the number of distinct elements is high.
Surprisingly, when the number of distinct elements is low (i.e., less than \SI{1000} in our experiments), ordered streams are faster than unordered ones, with a speedup factor up to~2x. 
This performance behavior is due to the synchronization cost of accessing a shared data structure (the \code{ConcurrentHashMap}) which does not pay off when the task-local \code{LinkedHashSet} merged with the fork-join algorithm are rather small.

CG and CGCC perform similarly on most queries.
However, the implementation of CGCC is more complex than CG, as described in Sec.~\ref{sec:codegen_options}.
CG simply uses the collector \code{groupingByConcurrent} (from the JCL) instead of \code{groupingBy}, leaving the user-defined collector not thread-safe, as it is accessed in a synchronized way, whereas CGCC requires the user to implement a thread-safe collector.

Q01 is the only TPC-H query with the following two orthogonal characteristics: 
(1)~the number of aggregated groups is small (leading to high contention in parallelized execution);
(2)~for each aggregated group, the aggregation is arithmetically expensive (many aggregated fields with non-trivial expressions).
To further study the impact of these aspects on parallelization, we implement two parametrized queries as microbenchmarks:
\emph{OneField} performs a grouped aggregation on a single field (Figure~\ref{fig:microquery-o3-1field}), whereas \emph{ManyFields} uses multiple fields (Figure~\ref{fig:microquery-o3-manyfields}).
Both queries perform a group-by operation on the result of a modulo operation over the column \code{id} containing consecutive integers.
The query parameter is the divisor of the modulo operation.
In this way, we have fine-grained control over the number of aggregated groups (i.e., the value of the divisor).

\begin{figure}
    \begin{verbatim}
SELECT SUM(p) FROM orders GROUP BY MOD(id, <M>)    
    \end{verbatim}
    \vspace{-5mm}
    \caption{OneField query (parametrized by <M>) for grouped aggregation with a single field, offering control over the number of aggregated groups (i.e., <M>).}
    \label{fig:microquery-o3-1field}
\end{figure}

\begin{figure}
    \begin{verbatim}
SELECT COUNT(*), SUM(p), AVG(p), MIN(p), MAX(p)
FROM orders GROUP BY MOD(id, <M>)    
    \end{verbatim}
    \vspace{-5mm}
    \caption{ManyFields query (parametrized by <M>) for grouped aggregation with multiple fields.}
    \label{fig:microquery-o3-manyfields}
\end{figure}

\begin{figure*}
    \centering

    \begin{subfigure}[b]{\textwidth}
        \centering
        \includegraphics[width=.4\linewidth]{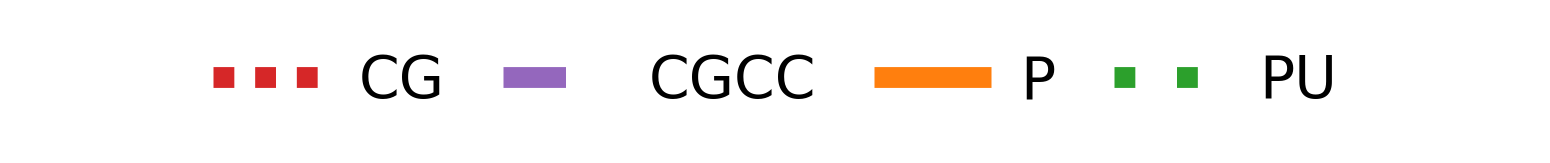}
    \end{subfigure}
    
    \begin{subfigure}[b]{0.47\textwidth}
        \centering
        \includegraphics[width=0.49\linewidth]{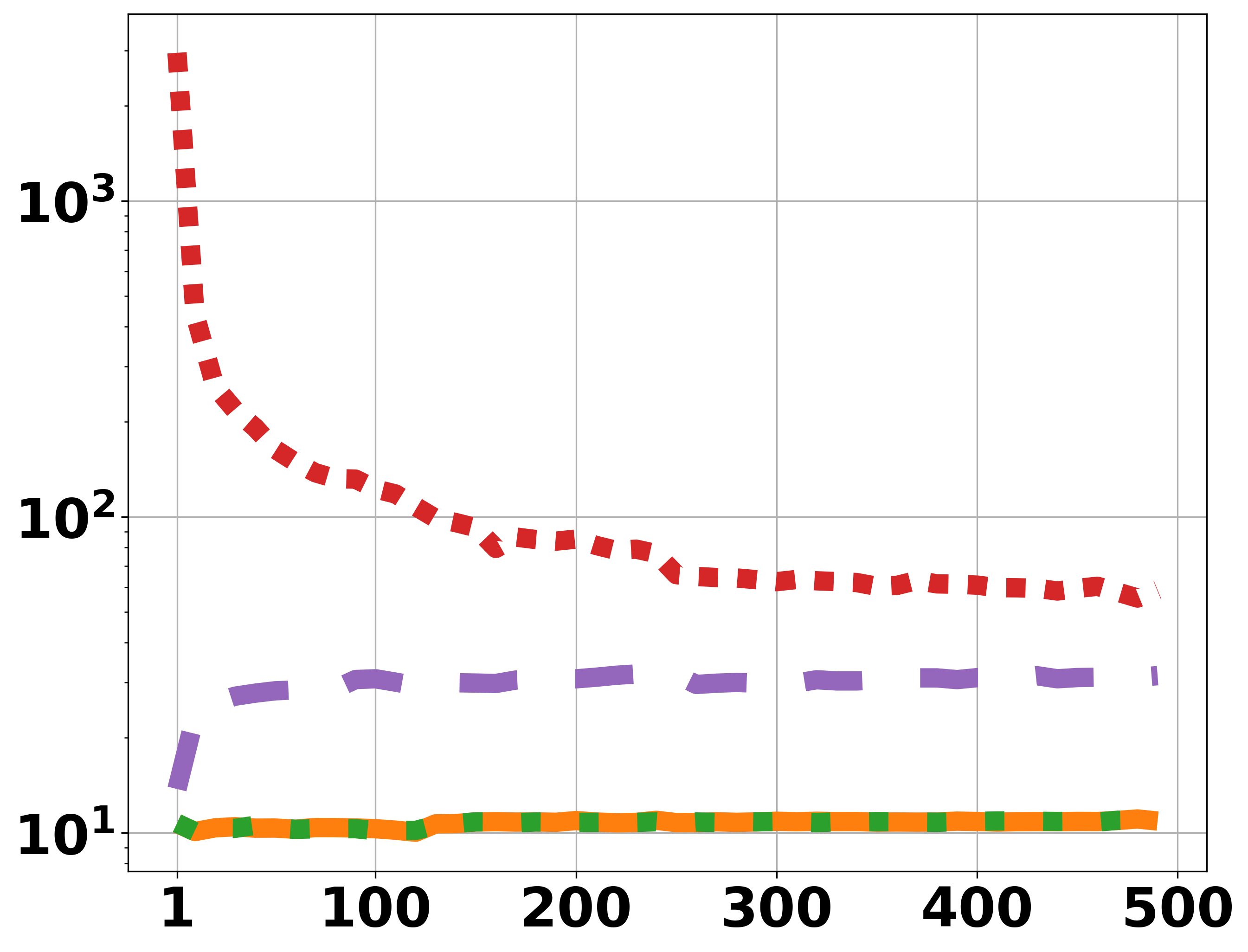}
        \includegraphics[width=0.49\linewidth]{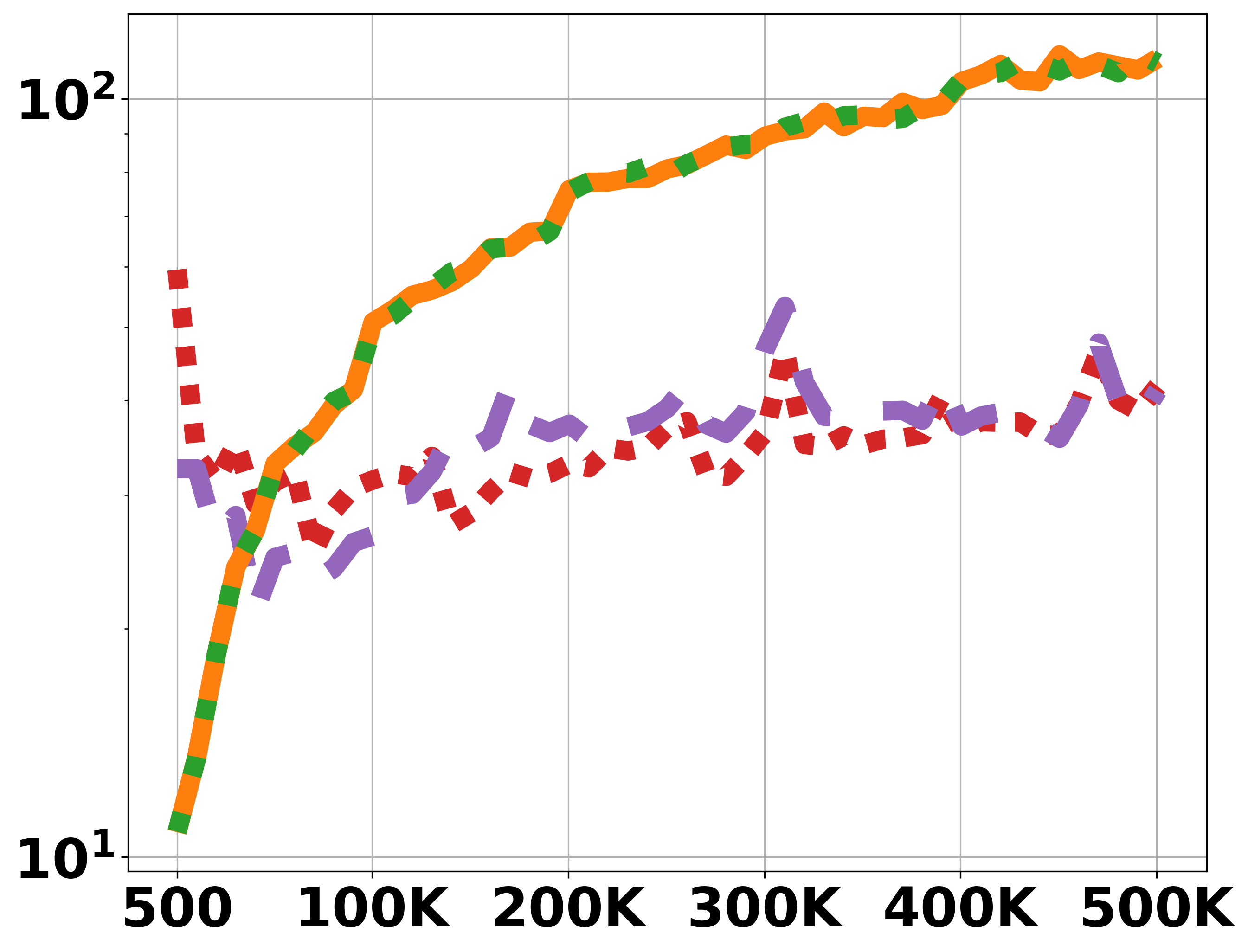}
        \caption{Intel - OneField}
    \end{subfigure}
    \begin{subfigure}[b]{0.02\textwidth}
        \centering
        \rule{0.3mm}{38mm}
    \end{subfigure}
    \begin{subfigure}[b]{0.47\textwidth}
        \centering
        \includegraphics[width=0.49\linewidth]{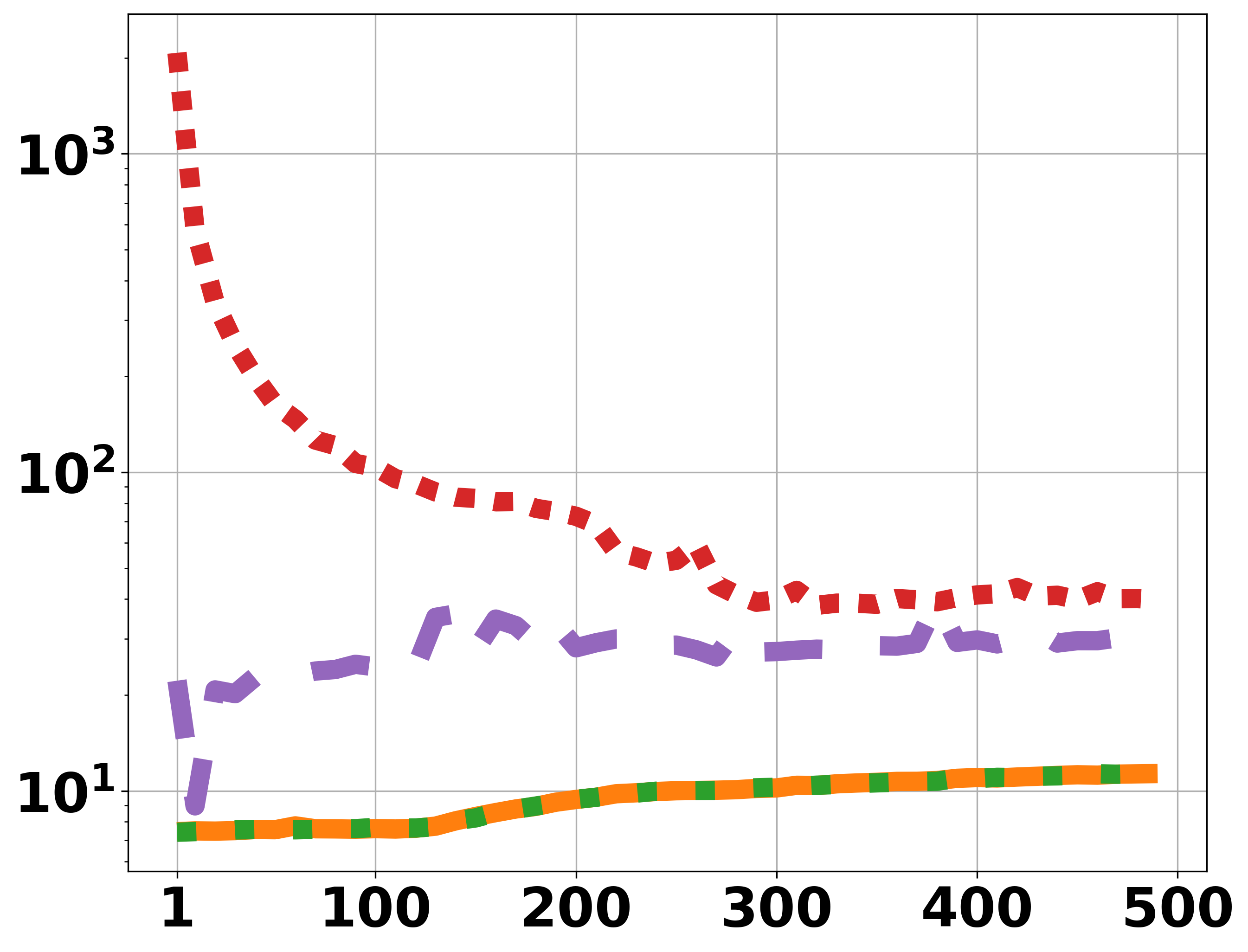}
        \includegraphics[width=0.49\linewidth]{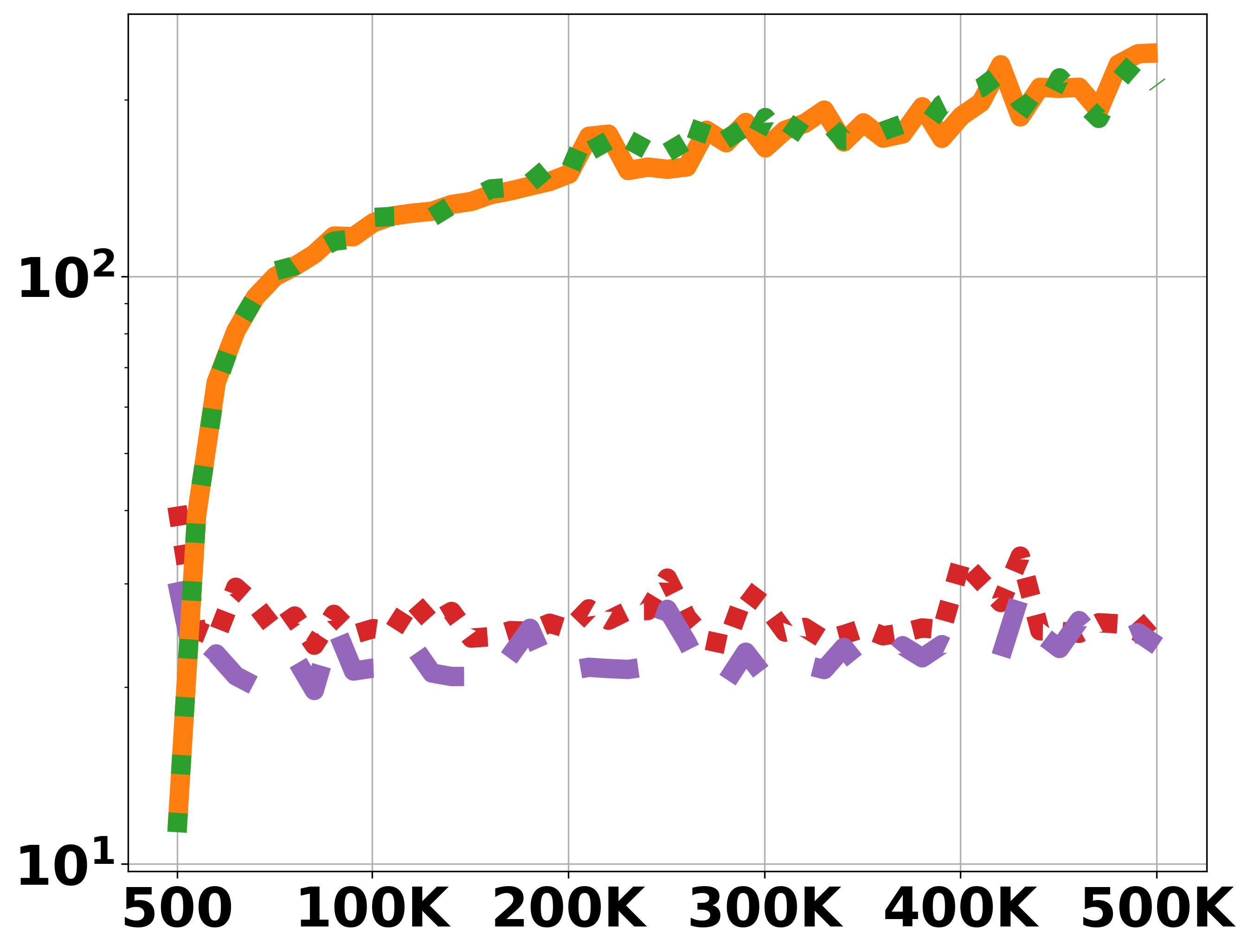}
        \caption{ARM - OneField}
    \end{subfigure}

    \begin{subfigure}[b]{0.47\textwidth}
        \centering
        \includegraphics[width=0.49\linewidth]{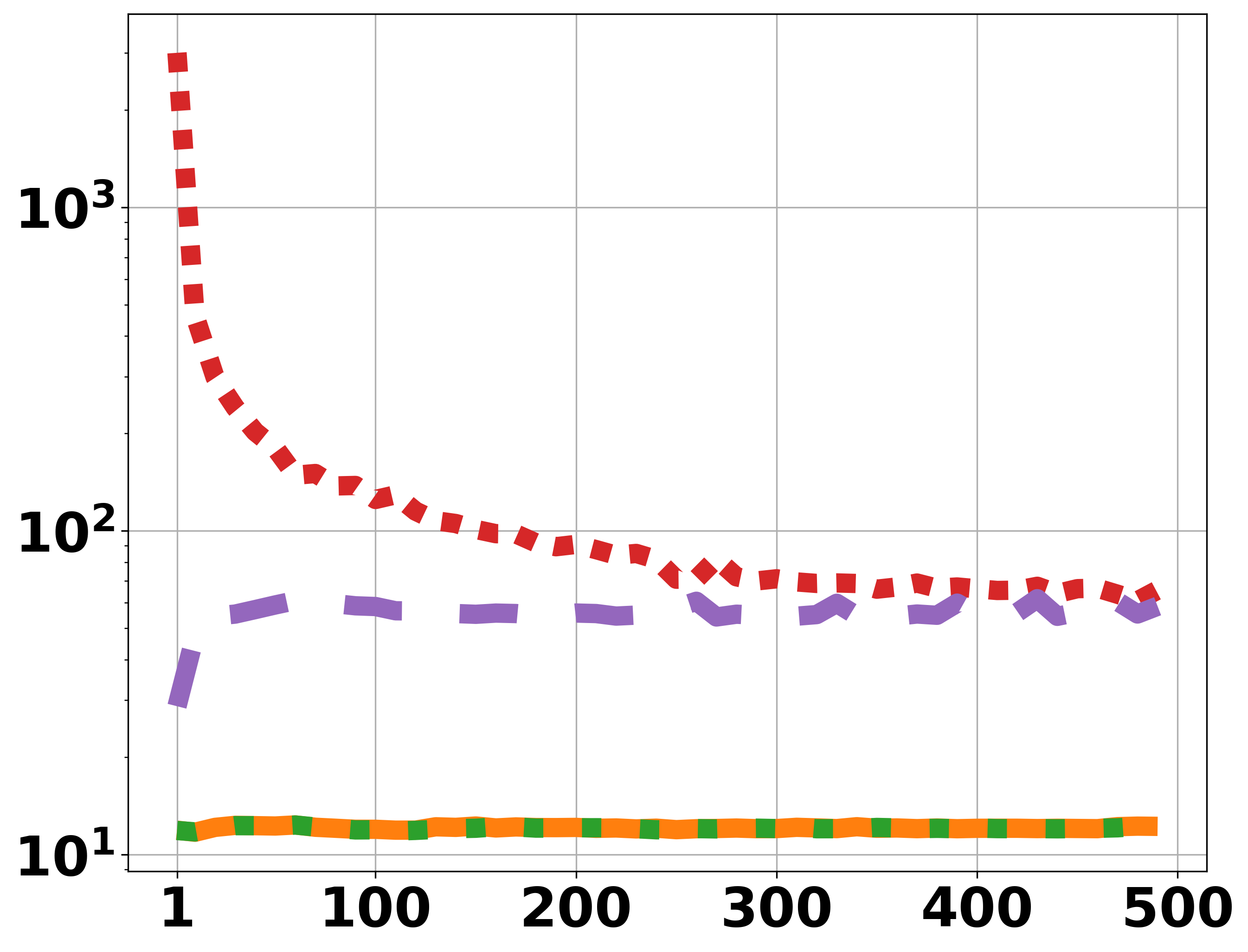}
        \includegraphics[width=0.49\linewidth]{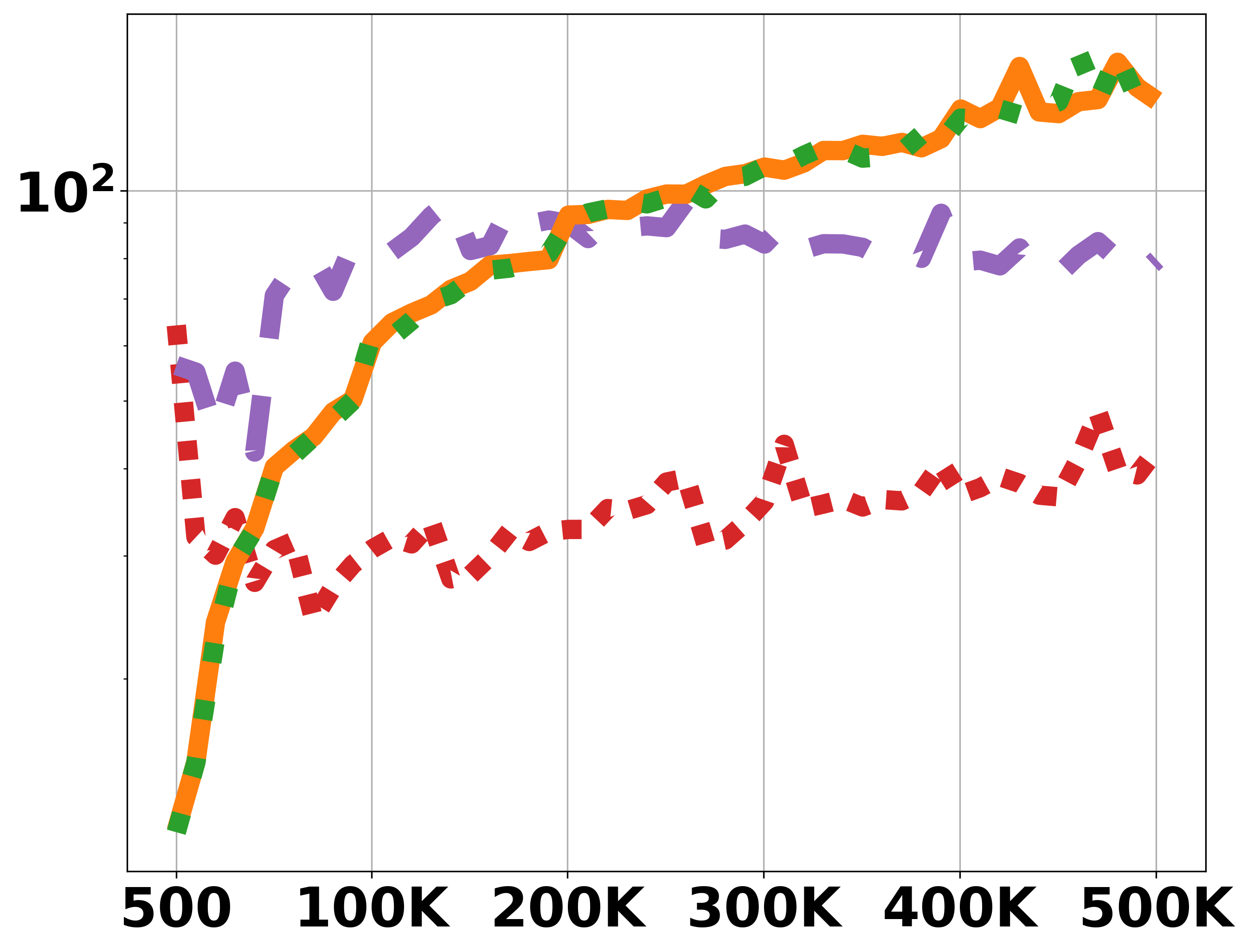}
        \caption{Intel - ManyFields}
    \end{subfigure}
    \begin{subfigure}[b]{0.02\textwidth}
        \centering
        \rule{0.3mm}{40mm}
    \end{subfigure}
    \begin{subfigure}[b]{0.47\textwidth}
        \centering
        \includegraphics[width=0.49\linewidth]{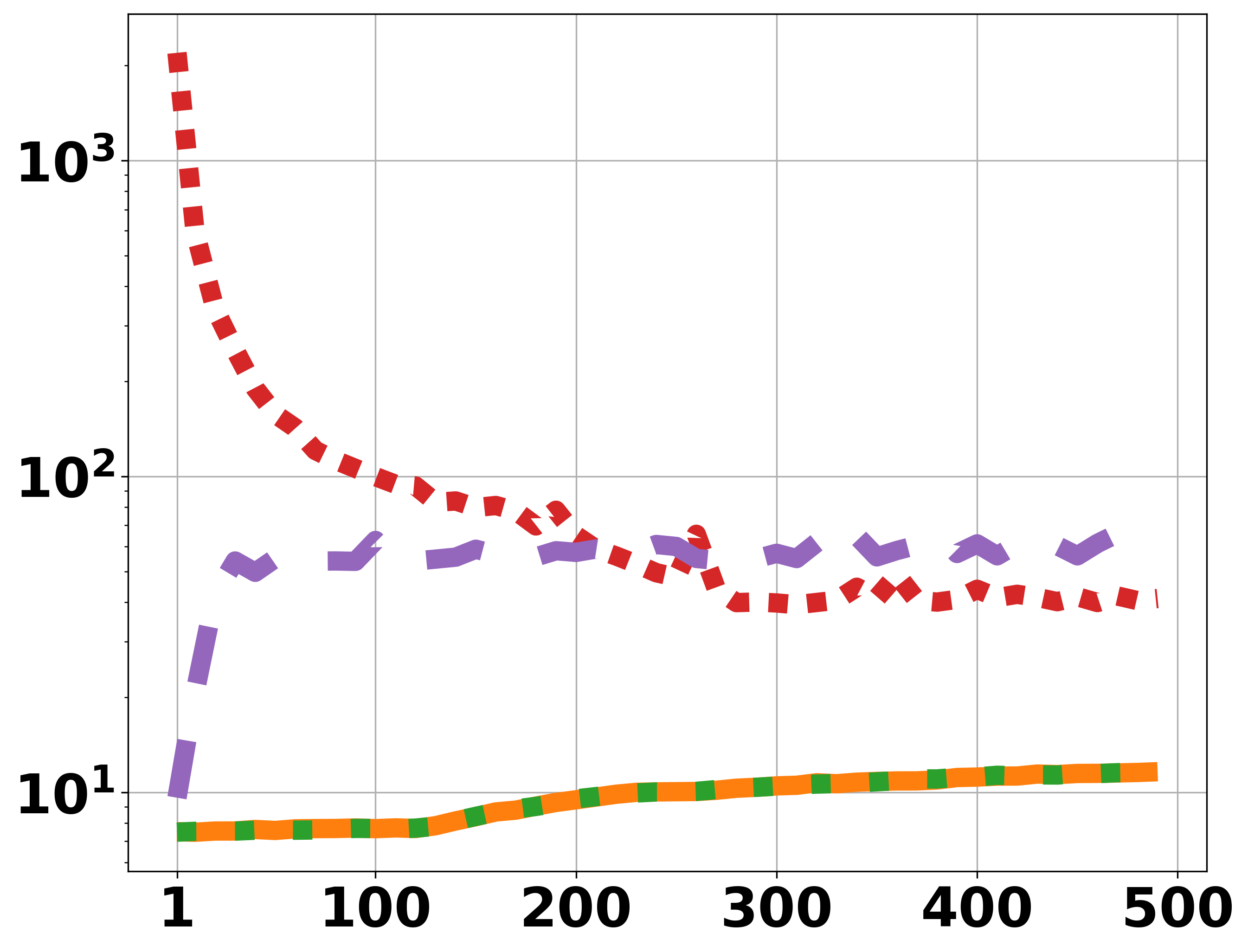}
        \includegraphics[width=0.49\linewidth]{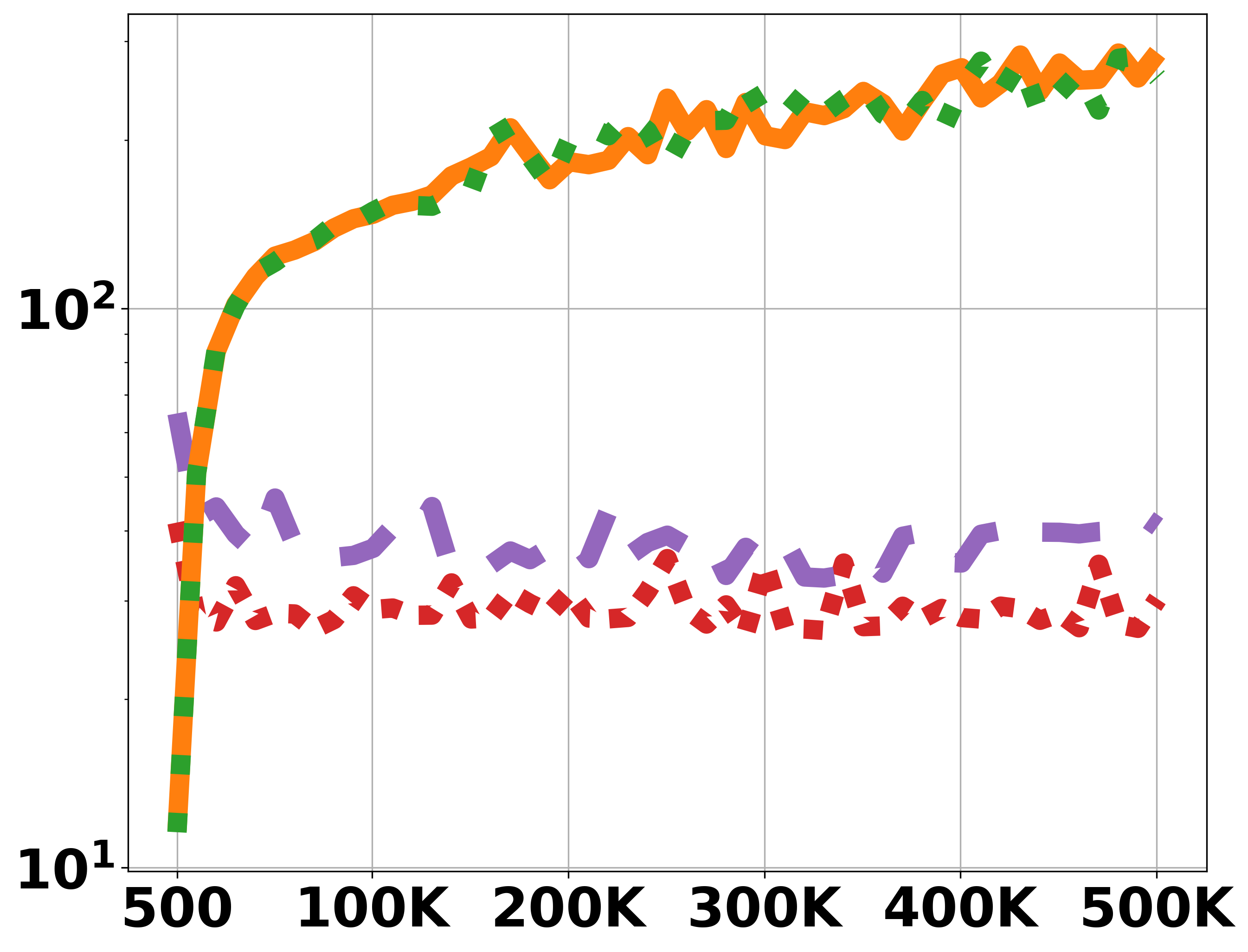}
        \caption{ARM - ManyFields}
    \end{subfigure}

    \caption{Execution time [ms] (y-axis) of the queries in Figure~\ref{fig:microquery-o3-1field} (OneField) and Figure~\ref{fig:microquery-o3-manyfields} (ManyFields) on JDK for different values of the modulo parameter, i.e., number of aggregated groups (x-axis).}
    \label{fig:vargroupsize}
\end{figure*}

For the parametrized queries, we evaluate the code generated for P, PU, CG, and CGCC.
The experiment results are depicted in Figure~\ref{fig:vargroupsize} for both Intel (left) and ARM (right).
The first row of charts corresponds to SingleField (Figure~\ref{fig:microquery-o3-1field}), the second row to ManyFields (Figure~\ref{fig:microquery-o3-manyfields}).
For each setting (SingleField or ManyFields, on Intel or ARM), we present two charts for different ranges of the modulo operator (1--500 vs.~500--500k), allowing us to observe different trends for small and large values of the operator, respectively. 

When the number of aggregated groups is very small, CG is much slower than PU and CGCC. 
CG uses lock-based synchronization for accessing the shared state (i.e., the aggregated fields within a single group); the smaller the number of groups, the more contended get the locks. 
In this range, P and PU are the fastest strategy by far, because contention is significant also for CGCC (even though less pronounced, thanks to the lock-free synchronization using atomic operations), whereas PU is completely free of contention.
However, when the number of aggregated groups increases, the trends are inverted, i.e., PU gets slower and CG faster.
There are two orthogonal aspects that contribute to these trends. 
First, P and PU create more thread-local maps (one for each executed task), all these maps are resized multiple times, which is an expensive operation. 
These thread-local maps also need to be merged; with an increasing number of groups, they get larger and merging them has high cost (which is, on the other hand, negligible when the maps are small). 
Moreover, the atomic operations introduce a slowdown compared to the same operations on primitive fields. 
Since acquiring an uncontended lock is rather cheap on the JVM~\cite{goetz2006JCP}, CG is even faster than CGCC, which we consider a surprising result.
This is particularly evident on ManyFields (second row of charts in Figure~\ref{fig:vargroupsize}), since once the lock is acquired by a thread, it can update multiple values without additional contention, in contrast to CGCC.

\textbf{Answer to RQ2:}
Our analysis shows that there is no strategy that always wins. 
Regarding aggregations, our analysis suggests that developers should use the P/PU strategies only when performing scalar aggregations (i.e, non-grouped ones) or when the expected number or groups is small. 
In these cases, CG and CGCC may suffer from high contention.
On the other hand, for grouped aggregations resulting in a large number of groups, we consider CG the best parallelization strategy. 
As the number of aggregated groups increases, the probability of contenting the access to the same object by multiple threads decreases.
Moreover, given the additional complexity required for implementing CGCC, and considering that it does not consistently yield better performance, we believe that CGCC is not worth the extra implementation effort.

However, we note that CGCC has an important benefit compared to CG.
As the microbenchmark results show, the performance of CGCC does not degrade in an unacceptable way for very small aggregated groups.
Therefore, CGCC can still be considered a viable strategy when it is impossible to predict the order of magnitude of the number of aggregated groups. 

Regarding the distinct operation, our analysis suggests that developers should generally use the PU strategy, as mentioned in the Java documentation~\cite{javadoc-distinct}. However, surprisingly, our analysis shows that if the number of distinct elements in the stream is small, the P strategy outperforms PU with a speedup up to~2x.

\subsection{Stream vs.~Imperative (RQ3)}\label{sec:eval:stream_imperative}

\begin{table*}[t]
    \setlength{\tabcolsep}{4pt}
    \centering
    \begin{minipage}{.48\textwidth}
        \centering
        \begin{tabular}{lr|rr||r|rr}
\toprule
& \multicolumn{3}{c||}{\textbf{JDK, Intel}} & \multicolumn{3}{c}{\textbf{JDK, ARM}} \\ 
\midrule
& \textbf{Stream} & \multicolumn{2}{c||}{\textbf{Imperative}} & \textbf{Stream} & \multicolumn{2}{c}{\textbf{Imperative}} \\ 
\midrule
\textbf{Q01} & 365$\pm$4 & 346$\pm$4 & \textbf{1.06x} & 613$\pm$3 & 560$\pm$6 & \textbf{1.09x} \\ 
\textbf{Q02} & 76$\pm$0 & 67$\pm$0 & \textbf{1.13x} & 141$\pm$1 & 123$\pm$0 & \textbf{1.15x} \\ 
\textbf{Q03} & 321$\pm$3 & 239$\pm$1 & \textbf{1.34x} & 627$\pm$2 & 436$\pm$15 & \textbf{1.44x} \\ 
\textbf{Q04} & 392$\pm$3 & 315$\pm$5 & \textbf{1.25x} & 616$\pm$56 & 518$\pm$31 & \textbf{1.19x} \\ 
\textbf{Q05} & 311$\pm$3 & 224$\pm$12 & \textbf{1.39x} & 666$\pm$4 & 417$\pm$3 & \textbf{1.60x} \\ 
\textbf{Q06} & 120$\pm$1 & 95$\pm$1 & \textbf{1.26x} & 199$\pm$0 & 129$\pm$2 & \textbf{1.55x} \\ 
\textbf{Q07} & 293$\pm$0 & 215$\pm$2 & \textbf{1.37x} & 582$\pm$5 & 376$\pm$6 & \textbf{1.55x} \\ 
\textbf{Q08} & 367$\pm$1 & 273$\pm$2 & \textbf{1.34x} & 705$\pm$21 & 470$\pm$10 & \textbf{1.50x} \\ 
\textbf{Q09} & 808$\pm$12 & 699$\pm$14 & \textbf{1.16x} & 1259$\pm$13 & 1090$\pm$8 & \textbf{1.15x} \\ 
\textbf{Q10} & 355$\pm$1 & 263$\pm$1 & \textbf{1.35x} & 558$\pm$6 & 385$\pm$9 & \textbf{1.45x} \\ 
\textbf{Q11} & 46$\pm$0 & 47$\pm$2 & \textbf{0.98x} & 92$\pm$2 & 83$\pm$0 & \textbf{1.11x} \\ 
\textbf{Q12} & 296$\pm$3 & 254$\pm$1 & \textbf{1.17x} & 491$\pm$8 & 351$\pm$1 & \textbf{1.40x} \\ 
\textbf{Q13} & 429$\pm$6 & 343$\pm$7 & \textbf{1.25x} & 682$\pm$8 & 593$\pm$25 & \textbf{1.15x} \\ 
\textbf{Q14} & 178$\pm$2 & 105$\pm$1 & \textbf{1.70x} & 386$\pm$8 & 181$\pm$1 & \textbf{2.14x} \\ 
\textbf{Q15} & 331$\pm$5 & 222$\pm$1 & \textbf{1.49x} & 824$\pm$74 & 316$\pm$6 & \textbf{2.61x} \\ 
\textbf{Q16} & 102$\pm$0 & 96$\pm$1 & \textbf{1.06x} & 155$\pm$2 & 138$\pm$3 & \textbf{1.12x} \\ 
\textbf{Q17} & 844$\pm$16 & 628$\pm$6 & \textbf{1.34x} & 1301$\pm$53 & 985$\pm$17 & \textbf{1.32x} \\ 
\textbf{Q18} & 652$\pm$7 & 464$\pm$10 & \textbf{1.41x} & 1205$\pm$22 & 874$\pm$8 & \textbf{1.38x} \\ 
\textbf{Q19} & 275$\pm$2 & 206$\pm$11 & \textbf{1.33x} & 473$\pm$4 & 385$\pm$2 & \textbf{1.23x} \\ 
\textbf{Q20} & 546$\pm$3 & 383$\pm$2 & \textbf{1.43x} & 850$\pm$11 & 560$\pm$11 & \textbf{1.52x} \\ 
\textbf{Q21} & 2240$\pm$32 & 1788$\pm$18 & \textbf{1.25x} & 3921$\pm$30 & 2768$\pm$48 & \textbf{1.42x} \\ 
\textbf{Q22} & 174$\pm$3 & 116$\pm$1 & \textbf{1.49x} & 271$\pm$4 & 202$\pm$3 & \textbf{1.34x} \\ 
 \hline 
 \multicolumn{2}{l|}{\textbf{Geo Mean}} & & \textbf{1.29x} & & & \textbf{1.39x} \\ 
\bottomrule
\end{tabular}
    \end{minipage}%
    \hfill
    \begin{minipage}{.48\textwidth}
        \centering
        \begin{tabular}{lr|rr||r|rr}
\toprule
& \multicolumn{3}{c||}{\textbf{GraalVM, Intel}} & \multicolumn{3}{c}{\textbf{GraalVM, ARM}} \\ 
\midrule
& \textbf{Stream} & \multicolumn{2}{c||}{\textbf{Imperative}} & \textbf{Stream} & \multicolumn{2}{c}{\textbf{Imperative}} \\ 
\midrule
\textbf{Q01} & 328$\pm$4 & 307$\pm$4 & \textbf{1.07x} & 550$\pm$9 & 520$\pm$4 & \textbf{1.06x} \\ 
\textbf{Q02} & 67$\pm$1 & 62$\pm$0 & \textbf{1.08x} & 132$\pm$4 & 120$\pm$4 & \textbf{1.10x} \\ 
\textbf{Q03} & 311$\pm$2 & 208$\pm$2 & \textbf{1.49x} & 634$\pm$25 & 377$\pm$18 & \textbf{1.68x} \\ 
\textbf{Q04} & 333$\pm$4 & 256$\pm$6 & \textbf{1.30x} & 644$\pm$7 & 409$\pm$4 & \textbf{1.57x} \\ 
\textbf{Q05} & 247$\pm$32 & 189$\pm$1 & \textbf{1.31x} & 490$\pm$13 & 416$\pm$31 & \textbf{1.18x} \\ 
\textbf{Q06} & 139$\pm$0 & 98$\pm$0 & \textbf{1.42x} & 219$\pm$3 & 152$\pm$4 & \textbf{1.44x} \\ 
\textbf{Q07} & 234$\pm$2 & 195$\pm$2 & \textbf{1.20x} & 507$\pm$10 & 341$\pm$9 & \textbf{1.49x} \\ 
\textbf{Q08} & 294$\pm$3 & 226$\pm$0 & \textbf{1.30x} & 512$\pm$10 & 477$\pm$9 & \textbf{1.07x} \\ 
\textbf{Q09} & 429$\pm$10 & 360$\pm$2 & \textbf{1.19x} & 954$\pm$60 & 846$\pm$58 & \textbf{1.13x} \\ 
\textbf{Q10} & 304$\pm$4 & 243$\pm$1 & \textbf{1.25x} & 528$\pm$8 & 422$\pm$12 & \textbf{1.25x} \\ 
\textbf{Q11} & 38$\pm$0 & 39$\pm$0 & \textbf{0.95x} & 84$\pm$1 & 74$\pm$0 & \textbf{1.12x} \\ 
\textbf{Q12} & 268$\pm$4 & 245$\pm$2 & \textbf{1.09x} & 519$\pm$5 & 444$\pm$2 & \textbf{1.17x} \\ 
\textbf{Q13} & 289$\pm$6 & 246$\pm$8 & \textbf{1.17x} & 572$\pm$5 & 411$\pm$6 & \textbf{1.39x} \\ 
\textbf{Q14} & 154$\pm$0 & 113$\pm$5 & \textbf{1.37x} & 268$\pm$4 & 156$\pm$0 & \textbf{1.71x} \\ 
\textbf{Q15} & 292$\pm$2 & 175$\pm$0 & \textbf{1.67x} & 573$\pm$2 & 259$\pm$4 & \textbf{2.21x} \\ 
\textbf{Q16} & 85$\pm$1 & 78$\pm$1 & \textbf{1.09x} & 147$\pm$2 & 123$\pm$2 & \textbf{1.19x} \\ 
\textbf{Q17} & 589$\pm$11 & 381$\pm$3 & \textbf{1.54x} & 988$\pm$6 & 553$\pm$15 & \textbf{1.78x} \\ 
\textbf{Q18} & 482$\pm$6 & 346$\pm$8 & \textbf{1.40x} & 1156$\pm$12 & 642$\pm$29 & \textbf{1.80x} \\ 
\textbf{Q19} & 249$\pm$3 & 210$\pm$4 & \textbf{1.18x} & 492$\pm$5 & 386$\pm$3 & \textbf{1.28x} \\ 
\textbf{Q20} & 346$\pm$6 & 248$\pm$2 & \textbf{1.40x} & 682$\pm$24 & 436$\pm$65 & \textbf{1.56x} \\ 
\textbf{Q21} & 1944$\pm$42 & 1407$\pm$49 & \textbf{1.38x} & 3647$\pm$45 & 2286$\pm$36 & \textbf{1.60x} \\ 
\textbf{Q22} & 125$\pm$1 & 76$\pm$1 & \textbf{1.64x} & 231$\pm$4 & 121$\pm$1 & \textbf{1.91x} \\ 
 \hline 
 \multicolumn{2}{l|}{\textbf{Geo Mean}} & & \textbf{1.28x} & & & \textbf{1.41x} \\ 
\bottomrule
\end{tabular}
    \end{minipage}
    
    \caption{Execution time [ms] on Intel and ARM with margin of error (half of the confidence interval) and speedup factor of the imperative version of the JEDI queries on JDK (left) and GraalVM (right). Baseline: sequential streams (O1+O2). SF$=$1.}
    \label{tab:vsimpertative}
\end{table*}

We address RQ3 extending our code-generator backend by integrating a conversion from SQL queries into loop-based, imperative Java code.
This implementation is inspired by the data-centric model~\cite{hyper}, later refined in the data-centric evaluation with callbacks~\cite{lb2}.
Thanks to this approach, we can compare the performance of the stream implementation and the imperative implementation, evaluating the so-called ``overhead of the Stream API'', i.e.,  the performance cost associated with the abstractions required to use a declarative way of expressing computations.
Remarkably, this approach results in a very fair comparison, since we can compare semantically equivalent implementations that differ only in the way of expressing the computation, while all other implementation aspects, such as the order of operations as well as all the data-structures involved in the computation, are exactly the same.

Table~\ref{tab:vsimpertative} compares the data of the imperative and the stream-based implementation on JDK (left) and GraalVM (right).
We show the query execution time of best stream implementation (O1+O2) and of the imperative implementation, as well as the speedup factor achieved by the latter. 
As the table show, the imperative implementation is faster than the stream-based implementation on most benchmarks.
On Intel, the imperative implementations show speedup factors ranging from 0.98x (Q11) to 1.70x (Q14) on JDK and from 0.95x (Q11) to 1.67x (Q15) on GraalVM. 
On JDK the geometric mean of the speedup is 1.29x (Intel) and 1.39x (ARM), while on GraalVM it is 1.28x (Intel) and 1.41x (ARM).
While GraalVM is usually faster than JDK, the speedup factors are similar.

\textbf{Answer to RQ3:}
Our analysis shows that the imperative code is faster than stream-based code, with a geometric mean of the speedup ranging from 1.28x (Intel, JDK) to 1.41x (ARM, GraalVM).
While this is expected, we note that related work reported higher speedups for imperative code~\cite{streamliner,parstreamliner}.
We believe that the performance of the Stream API implementation has improved over time.
Nonetheless, we note that for performance-critical code, imperative code is still substantially faster.

\subsection{Source-code Complexity Metrics (RQ4)}\label{sec:eval:complexity_metrics}

In this section, we answer to RQ4.
Recent research~\cite{stream_rct} using a randomized control trial show that developers implement solutions for the same tasks faster and more correct when using streams instead of loops-based code. 
Thanks to JEDI, we can complement this study, comparing code complexity metrics of the stream-based and imperative query implementations. 
Similarly to Sec.~\ref{sec:eval:stream_imperative}, we can perform a fair comparison thanks to automatic code generation. 

For this experiment, we report cognitive complexity~\cite{cognitive-complexity,cognitive-complexity-2}, design complexity~\cite{design-complexity}, and cyclomatic complexity~\cite{cyclomatic-complexity}.
Table~\ref{tab:complexity-metric} shows the metrics calculated with IntelliJ~\cite{intellij-idea} on all the methods generated in JEDI by converting the TPC-H queries.

\textbf{Answer to RQ4:}
The metrics, particularly cognitive complexity, demonstrate that code using streams is easier to understand than imperative code. We confirm the finding of~\cite{stream_rct}, since declarative code is considered less error-prone and easier to maintain than imperative code.
Cognitive complexity of baseline streams (i.e., no option enabled) is smaller than the imperative version by a factor of 2.43x and by a factor of 1.62x for optimized streams (i.e., enabling both O1 and O2).
It is worth noticing that enabling O1+O2 increases the complexity metrics compared to the baseline stream implementation, also because O1 results in more complex boolean expressions. 
However, even with O1+O2, the complexity metrics are still lower than for imperative code.

\begin{table}[]

\setlength{\tabcolsep}{4pt}

\begin{tabular}{cccc}
\multicolumn{1}{l}{}                                              & \multicolumn{1}{c}{\textbf{\begin{tabular}[c]{@{}c@{}}Cognitive\\ Complexity\end{tabular}}} & \multicolumn{1}{c}{\textbf{\begin{tabular}[c]{@{}c@{}}Design\\ Complexity\end{tabular}}} & \multicolumn{1}{c}{\textbf{\begin{tabular}[c]{@{}c@{}}Cyclomatic\\ Complexity\end{tabular}}} \\ \hline
\textbf{Stream}                                                   & 318                                                                                         & 1350                                                                                     & 1415                                                                                         \\
\textbf{\begin{tabular}[c]{@{}c@{}}Stream (O1+O2)\end{tabular}} & 476                                                                                         & 1504                                                                                     & 1572                                                                                         \\
\textbf{Imperative}                                               & 772                                                                                         & 1566                                                                                     & 1636                                                                                         \\ \hline
\end{tabular}
\caption{Complexity metrics of the JEDI suite (all queries): stream (no option enabled); stream (O1+O2); imperative.}
\label{tab:complexity-metric}
\end{table}

\section{Threats to Validity}\label{sec:validity}

\textbf{JEDI evaluates only streams generated by converting SQL queries, which may not cover all possible usage patterns.} 
The proposed extension of S2S does not cover all the methods offered by the Stream API.
However, we note that SQL is a data-processing language and the Stream API is designed with data processing in mind, therefore the kind of computations match.
Indeed, existing studies on the usage of the Stream API in real-world applications~\cite{on-the-wild-khatchadourian,on-the-wild-costa,on-the-wild-eduardo} report that real usages  match many patterns that we generate with our S2S extension.
We report the following usage patterns that are covered by our code generator:
\begin{enumerate}
    
    \item The most used terminal operator is \code{collect(...)}~\cite{on-the-wild-costa,on-the-wild-eduardo}, which is the most common terminal operator in our work.
    
    \item Most of the collectors return lists, sets, maps, and strings~\cite{on-the-wild-khatchadourian,on-the-wild-costa,on-the-wild-eduardo}. Our work does not cover the latter case, but it extensively covers streams returning lists (each main \code{exec} method returns a list), sets (the implementations of semi and anti joins), and maps (the implementations of both group-by and hash-join operations).
    
    \item Mapping and filtering are by far the most popular intermediate operations~\cite{on-the-wild-khatchadourian,on-the-wild-costa,on-the-wild-eduardo}. All the benchmarks generated in this work involve mapping and/or filtering elements.
\end{enumerate}

\textbf{The stream source is an important factor for stream performance, but our approach evaluates mostly arrays as data source.}
While we mostly evaluate arrays as a stream data source (using \code{Arrays.toStream(...)}), it is important to note that we also compare them with imperative code that uses loops over arrays.
Our approach can be extended to use other collections as data sources, e.g., lists or sets, but for fairness, we believe that in this case, the imperative code should use loops over those collections, too.
In this case, we expect that the performance difference between the stream-from-collection code and the loop-over-collection code would be in line with the stream-from-array code compared to the loop-over-array code that we generate.
Moreover, we note that the most common stream source is \code{ArrayList}~\cite{on-the-wild-eduardo}, which is backed by an array, and the implementations of the spliterators of array-based streams and \code{ArrayList}-based streams are very similar.

\textbf{Incomplete coverage of the Stream API.}
Our code generation covers many methods of the Stream API.
However, there are a few methods that are not exercised in JEDI.
Specifically, the intermediate stream operation \code{Stream.skip(int n)}, which allows skipping a given number of elements, is used by our generator for the SQL operator \code{OFFSET}, but this operator is not used in TPC-H.
A few stream operations, such as \code{takeWhile}, \code{dropWhile}, and \code{iterate}, are inherently difficult to cover in our approach, because there are no counterparts of these operations in SQL.

\textbf{An important contribution of this work is to analyze the performance of different parallelization strategies, but performance of parallel execution is machine-dependent.}
Our evaluation of the parallelization strategies (Sec.~\ref{sec:eval:stream_parallel}) makes use of two very different CPUs, namely an Intel Xeon Gold~6326 (16 cores) and an ARM Neoverse-N1 (80 cores). 
While parallelization performance is always machine-dependent, our results show similar trends on both machines.

\textbf{Coarse-grained granularity of parallelization options.}
When our code generator applies a certain option, all the stream pipelines generated from the input query will use that option.
While this approach is fine for the sequential options described in Sec.~\ref{sec:codegen_options} (as they generally lead to better performance), the proposed parallel strategies would benefit from more flexibility.
Currently, all streams generated for a query are parallelized with the same option. 
However, for each individual stream, a different parallelization option (or sequential execution) may yield better performance. 
\section{Related Work}\label{sec:relwork}

\textbf{Benchmarking the Java Stream API.}
While there are many benchmark suites for the JVM~\cite{dacapo,dacapo-chopin,specjvm98,specjvm2008,scalabench,renaissance}, none of them explicitly target the Stream API, and most of them do not use it at all.
Renaissance~\cite{renaissance} contains a few benchmarks that make use of streams.
In contrast, our work explicitly targets the Stream API---we generate many benchmarks based on the Stream API and we compare the performance of multiple, semantically equivalent implementations. 
In this way, we derive suggestions that can help developers get the best performance when using the Stream API. 

Our work uses the S2S~\cite{s2s} tool to convert SQL queries into Java code, and extends it with code-generation options an with an imperative backed. 
These features were not present in S2S.
Moreover, S2S was only used to generate a small benchmark suite (BSS) composed of 8~very simple queries, without evaluating the performance of the Stream API.
In contrast, our work uses the real-world TPC-H database benchmark to generate JEDI. 
TPC-H is far more complex and uses more operators than the example queries used in~\cite{s2s}; this required us to extend S2S to cover the missing SQL operations (e.g., left joins, semi joins, and anti joins). 
Therefore, JEDI covers many more methods and uses of the Stream API than BSS.

\textbf{Stream performance optimizations.}
Existing research~\cite{streamliner,parstreamliner} has proposed optimizations for the Stream API, showing high speedup reaching the performance of imperative code.
However, their evaluations were executed on simple stream pipelines and small datasets.
Our work can certainly be helpful in the context of optimizing the Stream API, because researcher and the developers of the JVM can use JEDI, as well as generate new benchmarks, to evaluate the benefits of their optimizations.

Another important line of research~\cite{khatchadourian-opt-stream-1,khatchadourian-opt-stream-2,khatchadourian-opt-stream-3,khatchadourian-opt-stream-4} proposes to automatically refactor streams to exploit parallelization more efficiently. 
This work focuses on automatically detecting when ordered streams can be converted to unordered ones in a semantics-preserving way, as well as moving from parallel to sequential streams in case the streams are ordered and the pipelines contain stateful intermediate operations (which may harm parallelization).
These work are complementary to our approach, since in our case, streams can always be safely reordered and parallelized: 
Our benchmarks are derived from SQL queries, and SQL does not require the implementer to respect the insertion order of the table rows upon query execution.

\textbf{Profilers and usage-analyses of the Java Stream API.}
A recent work~\cite{stream_rct} evaluates developer productivity when using the Stream API in comparison to equivalent imperative code, showing that using the Stream API results in code that is less error-prone and easier to understand and maintain.
This work is complementary with ours and partially motivates ours, since developer productivity benefits from the Stream API, our work shows how to improve performance when using the Stream API.

In the literature, a profiler for the Stream API~\cite{streamprof} has been proposed, as well as a profiler for the Java Fork-Join framework~\cite{fjprof}, which is used by the Stream API to implement parallellism.
These profilers, as well as static analyses, have been used to analyze the usage of the Stream API.
Many studies~\cite{on-the-wild-khatchadourian,on-the-wild-costa,on-the-wild-tanaka,on-the-wild-eduardo} explore existing open-source projects (e.g., on GitHub) to study how the Stream API is commonly used in real-world applications.
These studies are orthogonal and complementary to our work.
Remarkably, as highlighted in Sec.~\ref{sec:validity}, our approach creates benchmarks which make use of the Stream API in a way that matches the real-world patterns detected in the aforementioned related work.

\section{Conclusions}\label{sec:conclusion}

In this paper we introduce JEDI, a benchmark suite for the Java Stream API.
We generate JEDI by automatically converting SQL queries.
Our conversion is parametric, since, given the same SQL query, it generates multiple code variants which use different operations.
In this way, we can analyze the performance of each variant.

With our analysis, we answer four important research questions.
First (RQ1), we suggest developers to fuse multiple calls to \code{filter} and to use \code{mapMulti} instead of \code{flatMap} for one-to-many transformations.
Second (RQ2), we suggest developers which parallelization strategy is more suitable
depending on data characteristics.
Third (RQ3), we find that imperative code generally outperforms stream-based implementations. 
We believe that 
JEDI can help the Java implementers in optimizing the Stream API to close this performance gap.
Fourth (RQ4), we report source-code complexity metrics for some of our sequential code variants.
The metrics show that the faster code variants suffer from an increased code complexity. 

As future work, we plan to extend JEDI to mitigate the threats to validity of our work.
First, we plan to extend the code generator to create identical streams with different data sources, comparing the performance of the different spliterators. Then, we plan to fully cover the Stream-API methods by creating handwritten benchmarks covering the methods that are not supported by our automatic conversion.
Finally, we plan to control the parallelization strategy for individual streams, rather than for whole queries.

\begin{acks}
This work has been supported by the Hasler Fundation (project 2025-03-13-355).
\end{acks}

\bibliographystyle{ACM-Reference-Format}
\bibliography{biblio}

@book{java8inaction,
  title={Java 8 in Action: Lambdas, Streams, and functional-style programming},
  author={Urma, Raoul-Gabriel and Fusco, Mario and Mycroft, Alan},
  year={2014},
  publisher={Manning Publications Co.}
}

@article{in-memory-db,
  title={In-memory databases: Challenges and opportunities from software and hardware perspectives},
  author={Tan, Kian-Lee and Cai, Qingchao and Ooi, Beng Chin and Wong, Weng-Fai and Yao, Chang and Zhang, Hao},
  journal={ACM Sigmod Record},
  volume={44},
  number={2},
  pages={35--40},
  year={2015},
  publisher={ACM New York, NY, USA}
}

@inproceedings{fjprof,
  author    = {Matteo Basso and
               Eduardo Rosales and
               Filippo Schiavio and
               Andrea Ros{\`{a}} and
               Walter Binder},
  title     = {Accurate Fork-Join Profiling on the Java Virtual Machine},
  booktitle = {Euro-Par 2022: Parallel Processing - 28th International Conference on Parallel and Distributed Computing},
  pages     = {35--50},
  publisher = {Springer},
  year      = {2022},
  url       = {https://doi.org/10.1007/978-3-031-12597-3\_3},
  doi       = {10.1007/978-3-031-12597-3\_3},
  bibsource = {dblp computer science bibliography, https://dblp.org}
}

@inproceedings{s2s,
  author    = {Filippo Schiavio and
               Andrea Ros{\`{a}} and
               Walter Binder},
  editor    = {Bernhard Scholz and
               Yukiyoshi Kameyama},
  title     = {{SQL} to Stream with {S2S:} An Automatic Benchmark Generator for the
               Java Stream {API}},
  booktitle = {Proceedings of the 21st {ACM} {SIGPLAN} International Conference on
               Generative Programming: Concepts and Experiences, {GPCE} 2022, Auckland,
               New Zealand, December 6-7, 2022},
  pages     = {179--186},
  publisher = {{ACM}},
  year      = {2022},
  url       = {https://doi.org/10.1145/3564719.3568699},
  doi       = {10.1145/3564719.3568699},
  bibsource = {dblp computer science bibliography, https://dblp.org}
}

@inproceedings{parstreamliner,
  author    = {Matteo Basso and
               Filippo Schiavio and
               Andrea Ros{\`{a}} and
               Walter Binder},
  title     = {Optimizing Parallel Java Streams},
  booktitle = {26th International Conference on Engineering of Complex Computer Systems,
               {ICECCS} 2022, Hiroshima, Japan, March 26-30, 2022},
  pages     = {23--32},
  publisher = {{IEEE}},
  year      = {2022},
  url       = {https://doi.org/10.1109/ICECCS54210.2022.00012},
  doi       = {10.1109/ICECCS54210.2022.00012},
  bibsource = {dblp computer science bibliography, https://dblp.org}
}

@article{dynq,
  author    = {Filippo Schiavio and
               Daniele Bonetta and
               Walter Binder},
  title     = {Language-Agnostic Integrated Queries in a Managed Polyglot Runtime},
  journal   = {Proc. {VLDB} Endow.},
  volume    = {14},
  number    = {8},
  pages     = {1414--1426},
  year      = {2021},
  url       = {http://www.vldb.org/pvldb/vol14/p1414-schiavio.pdf},
  doi       = {10.14778/3457390.3457405},
  bibsource = {dblp computer science bibliography, https://dblp.org}
}

@article{dynq_journal,
  author       = {Filippo Schiavio and
                  Daniele Bonetta and
                  Walter Binder},
  title        = {DynQ: a dynamic query engine with query-reuse capabilities embedded
                  in a polyglot runtime},
  journal      = {{VLDB} J.},
  volume       = {32},
  number       = {5},
  pages        = {1111--1135},
  year         = {2023},
  url          = {https://doi.org/10.1007/s00778-023-00784-2},
  doi          = {10.1007/S00778-023-00784-2},
  bibsource    = {dblp computer science bibliography, https://dblp.org}
}

@inproceedings{stream_rct,
  author       = {Nils Mehlhorn and
                  Stefan Hanenberg},
  title        = {Imperative versus Declarative Collection Processing: An {RCT} on the
                  Understandability of Traditional Loops versus the Stream {API} in
                  Java},
  booktitle    = {44th {IEEE/ACM} 44th International Conference on Software Engineering,
                  {ICSE} 2022, Pittsburgh, PA, USA, May 25-27, 2022},
  pages        = {1157--1168},
  publisher    = {{ACM}},
  year         = {2022},
  url          = {https://doi.org/10.1145/3510003.3519016},
  doi          = {10.1145/3510003.3519016},
  bibsource    = {dblp computer science bibliography, https://dblp.org}
}

@inproceedings{renaissance,
  author       = {Aleksandar Prokopec and
                  Andrea Ros{\`{a}} and
                  David Leopoldseder and
                  Gilles Duboscq and
                  Petr Tuma and
                  Martin Studener and
                  Lubom{\'{\i}}r Bulej and
                  Yudi Zheng and
                  Alex Villaz{\'{o}}n and
                  Doug Simon and
                  Thomas W{\"{u}}rthinger and
                  Walter Binder},
  title        = {Renaissance: Benchmarking Suite for Parallel Applications on the {JVM}},
  booktitle    = {Software Engineering 2020, Fachtagung des GI-Fachbereichs Softwaretechnik},
  series       = {{LNI}},
  volume       = {{P-300}},
  pages        = {145--146},
  publisher    = {Gesellschaft f{\"{u}}r Informatik e.V.},
  year         = {2020},
  url          = {https://doi.org/10.18420/SE2020\_44},
  doi          = {10.18420/SE2020\_44},
  bibsource    = {dblp computer science bibliography, https://dblp.org}
}

@article{streamliner,
	author = {M\o{}ller, Anders and Veileborg, Oskar Haarklou},
	title = {{Eliminating Abstraction Overhead of Java Stream Pipelines Using Ahead-of-Time Program Optimization}},
	year = {2020},
	volume = {4},
	number = {OOPSLA},
	journal = {Proc. ACM Program. Lang.},
	pages = {1--29}
}

@InProceedings{umbra,
  author    = {Thomas Neumann and Michael J. Freitag},
  booktitle = {{CIDR}},
  title     = {{Umbra: A Disk-Based System with In-Memory Performance}},
  year      = {2020},
}

@article{hyper,
author = {Neumann, Thomas},
title = {Efficiently Compiling Efficient Query Plans for Modern Hardware},
year = {2011},
publisher = {VLDB Endowment},
volume = {4},
number = {9},
journal = {Proc. VLDB Endow.},
pages = {539–550},
}

@InProceedings{dacapo,
  title={The {DaCapo} Benchmarks: {J}ava Benchmarking Development and Analysis},
  author={Blackburn, S. M. and Garner, R. and Hoffman, C. and Khan, A. M.
      and McKinley, K. S. and Bentzur, R. and Diwan, A. and Feinberg, D. 
      and Frampton, D. and Guyer, S. Z. and Hirzel, M. and Hosking, A.
      and Jump, M. and Lee, H. and Moss, J. E. B. and Phansalkar, A.
      and Stefanovi\'{c}, D. and {VanDrunen}, T. and von~Dincklage, D.
      and Wiedermann, B.},
  booktitle = {OOPSLA '06: Proceedings of the 21st annual ACM SIGPLAN conference on Object-Oriented Programing, Systems, Languages, and Applications},
  month = oct,
  year = {2006},
  pages = {169--190},
  location = {Portland, OR, USA},
  doi = {http://doi.acm.org/10.1145/1167473.1167488},
  publisher = {ACM Press},
  address = {New York, NY, USA},
}

@inproceedings{dacapo-chopin,
  author       = {Stephen M Blackburn and
                  Zixian Cai and
                  Rui Chen and
                  Xi Yang and
                  John Zhang and
                  John Zigman
                 },
  title        = {Rethinking {Java} Performance Analysis},
  booktitle    = {Proceedings of the 30th {ACM} International Conference on Architectural
                  Support for Programming Languages and Operating Systems, Volume 1,
                  {ASPLOS} 2025, Rotterdam, Netherlands, 30 March 2025 - 3 April 2025},
  publisher    = {{ACM}},
  year         = {2025},
  url          = {https://doi.org/10.1145/3669940.3707217},
  doi          = {10.1145/3669940.3707217},
}

@inproceedings{scalabench,
  title={Da capo con scala: Design and analysis of a scala benchmark suite for the java virtual machine},
  author={Sewe, Andreas and Mezini, Mira and Sarimbekov, Aibek and Binder, Walter},
  booktitle={Proceedings of the 2011 ACM international conference on Object oriented programming systems languages and applications},
  pages={657--676},
  year={2011}
}

@article{streamprof,
  author       = {Eduardo Rosales and
                  Matteo Basso and
                  Andrea Ros{\`{a}} and
                  Walter Binder},
  title        = {Profiling and Optimizing Java Streams},
  journal      = {Art Sci. Eng. Program.},
  volume       = {7},
  number       = {3},
  year         = {2023},
  url          = {https://doi.org/10.22152/programming-journal.org/2023/7/10},
  doi          = {10.22152/PROGRAMMING-JOURNAL.ORG/2023/7/10},
  bibsource    = {dblp computer science bibliography, https://dblp.org}
}

@article{zhang2015memory,
  title={In-memory big data management and processing: A survey},
  author={Zhang, Hao and Chen, Gang and Ooi, Beng Chin and Tan, Kian-Lee and Zhang, Meihui},
  journal={IEEE Transactions on Knowledge and Data Engineering},
  volume={27},
  number={7},
  pages={1920--1948},
  year={2015},
  publisher={IEEE}
}

@article{on-the-wild-eduardo,
  author       = {Eduardo Rosales and
                  Matteo Basso and
                  Andrea Ros{\`{a}} and
                  Walter Binder},
  title        = {Large-scale characterization of Java streams},
  journal      = {Softw. Pract. Exp.},
  volume       = {53},
  number       = {9},
  pages        = {1763--1792},
  year         = {2023},
  url          = {https://doi.org/10.1002/spe.3213},
  doi          = {10.1002/SPE.3213},
  bibsource    = {dblp computer science bibliography, https://dblp.org}
}

@InProceedings{on-the-wild-costa,
    author="Nostas, Joshua
    and Alcocer, Juan Pablo Sandoval
    and Costa, Diego Elias
    and Bergel, Alexandre",
    editor="Gervasi, Osvaldo
    and Murgante, Beniamino
    and Misra, Sanjay
    and Garau, Chiara
    and Ble{\v{c}}i{\'{c}}, Ivan
    and Taniar, David
    and Apduhan, Bernady O.
    and Rocha, Ana Maria A. C.
    and Tarantino, Eufemia
    and Torre, Carmelo Maria",
    title="How Do Developers Use the Java Stream API?",
    booktitle="Computational Science and Its Applications -- ICCSA 2021",
    year="2021",
    publisher="Springer International Publishing",
    address="Cham",
    pages="323--335",
    isbn="978-3-030-87007-2"
}

@inproceedings{on-the-wild-khatchadourian,
  author       = {Raffi Khatchadourian and
                  Yiming Tang and
                  Mehdi Bagherzadeh and
                  Baishakhi Ray},
  editor       = {Heike Wehrheim and
                  Jordi Cabot},
  title        = {An Empirical Study on the Use and Misuse of Java 8 Streams},
  booktitle    = {Fundamental Approaches to Software Engineering - 23rd International Conference, {FASE} 2020},
  series       = {Lecture Notes in Computer Science},
  volume       = {12076},
  pages        = {97--118},
  publisher    = {Springer},
  year         = {2020},
  url          = {https://doi.org/10.1007/978-3-030-45234-6\_5},
  doi          = {10.1007/978-3-030-45234-6\_5},
  bibsource    = {dblp computer science bibliography, https://dblp.org}
}

@article{on-the-wild-tanaka,
  title={A study on the current status of functional idioms in Java},
  author={Tanaka, Hiroto and Matsumoto, Shinsuke and Kusumoto, Shinji},
  journal={IEICE Transactions on Information and Systems},
  volume={102},
  number={12},
  pages={2414--2422},
  year={2019},
  publisher={The Institute of Electronics, Information and Communication Engineers}
}

@inproceedings{chokepoints1,
  title={TPC-H analyzed: Hidden messages and lessons learned from an influential benchmark},
  author={Boncz, Peter and Neumann, Thomas and Erling, Orri},
  booktitle={Technology Conference on Performance Evaluation and Benchmarking},
  pages={61--76},
  year={2013},
  organization={Springer}
}

@article{chokepoints2,
  title={Quantifying TPC-H choke points and their optimizations},
  author={Dreseler, Markus and Boissier, Martin and Rabl, Tilmann and Uflacker, Matthias},
  journal={Proceedings of the VLDB Endowment},
  volume={13},
  number={8},
  pages={1206--1220},
  year={2020},
  publisher={VLDB Endowment}
}

@inproceedings{khatchadourian-opt-stream-1,
  author       = {Raffi Khatchadourian and
                  Yiming Tang and
                  Mehdi Bagherzadeh and
                  Syed Ahmed},
  title        = {[Engineering Paper] {A} Tool for Optimizing Java 8 Stream Software
                  via Automated Refactoring},
  booktitle    = {18th {IEEE} International Working Conference on Source Code Analysis
                  and Manipulation, {SCAM} 2018, Madrid, Spain, September 23-24, 2018},
  pages        = {34--39},
  publisher    = {{IEEE} Computer Society},
  year         = {2018},
  url          = {https://doi.org/10.1109/SCAM.2018.00011},
  doi          = {10.1109/SCAM.2018.00011},
  bibsource    = {dblp computer science bibliography, https://dblp.org}
}

@inproceedings{khatchadourian-opt-stream-2,
  author       = {Yiming Tang and
                  Raffi Khatchadourian and
                  Mehdi Bagherzadeh and
                  Syed Ahmed},
  editor       = {Michel Chaudron and
                  Ivica Crnkovic and
                  Marsha Chechik and
                  Mark Harman},
  title        = {Towards safe refactoring for intelligent parallelization of Java 8
                  streams},
  booktitle    = {Proceedings of the 40th International Conference on Software Engineering:
                  Companion Proceeedings, {ICSE} 2018, Gothenburg, Sweden, May 27 -
                  June 03, 2018},
  pages        = {206--207},
  publisher    = {{ACM}},
  year         = {2018},
  url          = {https://doi.org/10.1145/3183440.3195098},
  doi          = {10.1145/3183440.3195098},
  bibsource    = {dblp computer science bibliography, https://dblp.org}
}

@inproceedings{khatchadourian-opt-stream-3,
  author       = {Raffi Khatchadourian and
                  Yiming Tang and
                  Mehdi Bagherzadeh and
                  Syed Ahmed},
  editor       = {Joanne M. Atlee and
                  Tevfik Bultan and
                  Jon Whittle},
  title        = {Safe automated refactoring for intelligent parallelization of Java
                  8 streams},
  booktitle    = {Proceedings of the 41st International Conference on Software Engineering,
                  {ICSE} 2019, Montreal, QC, Canada, May 25-31, 2019},
  pages        = {619--630},
  publisher    = {{IEEE} / {ACM}},
  year         = {2019},
  url          = {https://doi.org/10.1109/ICSE.2019.00072},
  doi          = {10.1109/ICSE.2019.00072},
  bibsource    = {dblp computer science bibliography, https://dblp.org}
}

@article{khatchadourian-opt-stream-4,
  author       = {Raffi Khatchadourian and
                  Yiming Tang and
                  Mehdi Bagherzadeh},
  title        = {Safe automated refactoring for intelligent parallelization of Java
                  8 streams},
  journal      = {Sci. Comput. Program.},
  volume       = {195},
  pages        = {102476},
  year         = {2020},
  url          = {https://doi.org/10.1016/j.scico.2020.102476},
  doi          = {10.1016/J.SCICO.2020.102476},
  bibsource    = {dblp computer science bibliography, https://dblp.org}
}

@inproceedings{duckdb,
  title={Duckdb: an embeddable analytical database},
  author={Raasveldt, Mark and M{\"u}hleisen, Hannes},
  booktitle={Proceedings of the 2019 international conference on management of data},
  pages={1981--1984},
  year={2019}
}

@inproceedings{cognitive-complexity,
author = {Campbell, G. Ann},
title = {Cognitive complexity: an overview and evaluation},
year = {2018},
isbn = {9781450357135},
publisher = {Association for Computing Machinery},
address = {New York, NY, USA},
url = {https://doi.org/10.1145/3194164.3194186},
doi = {10.1145/3194164.3194186},
booktitle = {Proceedings of the 2018 International Conference on Technical Debt},
pages = {57–58},
numpages = {2},
location = {Gothenburg, Sweden},
series = {TechDebt '18}
}

@article{cognitive-complexity-2,
title = {Cognitive complexity as a quantifier of version to version Java-based source code change: An empirical probe},
journal = {Information and Software Technology},
volume = {106},
pages = {31-48},
year = {2019},
issn = {0950-5849},
doi = {https://doi.org/10.1016/j.infsof.2018.09.002},
url = {https://www.sciencedirect.com/science/article/pii/S0950584918301903},
author = {Loveleen Kaur and Ashutosh Mishra},
}

@article{cyclomatic-complexity,
  title={A complexity measure},
  author={McCabe, Thomas J},
  journal={IEEE Transactions on software Engineering},
  number={4},
  pages={308--320},
  year={1976},
  publisher={IEEE}
}

@inproceedings{lb2,
  title={How to architect a query compiler, revisited},
  author={Tahboub, Ruby Y and Essertel, Gr{\'e}gory M and Rompf, Tiark},
  booktitle={Proceedings of the 2018 International Conference on Management of Data},
  pages={307--322},
  year={2018}
}

@article{jmh-do-dons-costa,
  title={What's wrong with my benchmark results? Studying bad practices in JMH benchmarks},
  author={Costa, Diego and Bezemer, Cor-Paul and Leitner, Philipp and Andrzejak, Artur},
  journal={IEEE Transactions on Software Engineering},
  volume={47},
  number={7},
  pages={1452--1467},
  year={2019},
  publisher={IEEE}
}

@inproceedings{graalvm,
  title={One VM to rule them all},
  author={W{\"u}rthinger, Thomas and Wimmer, Christian and W{\"o}{\ss}, Andreas and Stadler, Lukas and Duboscq, Gilles and Humer, Christian and Richards, Gregor and Simon, Doug and Wolczko, Mario},
  booktitle={Proceedings of the 2013 ACM international symposium on New ideas, new paradigms, and reflections on programming \& software},
  pages={187--204},
  year={2013}
}

@article{everything-vect-comp,
  title={Everything you always wanted to know about compiled and vectorized queries but were afraid to ask},
  author={Kersten, Timo and Leis, Viktor and Kemper, Alfons and Neumann, Thomas and Pavlo, Andrew and Boncz, Peter},
  journal={Proceedings of the VLDB Endowment},
  volume={11},
  number={13},
  pages={2209--2222},
  year={2018},
  publisher={VLDB Endowment}
}

@Book{junit,
  author    = {Petar Tahchiev and Felipe Leme and Vincent Massol and Gary Gregory},
  publisher = {Manning Publications Company},
  title     = {JUnit in Action, 2nd Edition},
  year      = {2011},
  doi       = {10.21019/9781582121994.ch9},
}

@book{goetz2006JCP,
  title={Java concurrency in practice},
  author={Goetz, Brian},
  year={2006},
  publisher={Pearson Education}
}

@Misc{tpch,
  author  = {{TPC}},
  title   = {{TPC-H - Homepage}},
  note    = {\url{http://www.tpc.org/tpch/}},
  urldate = {2024-01-31},
  year    = {2024},
}

@Misc{specjvm98,
  author  = {{SPEC}},
  title   = {{SpecJVM98}},
  note    = {\url{https://www.spec.org/jvm98/}},
  urldate = {2024-01-31},
  year    = {2008},
}

@Misc{specjvm2008,
  author  = {{SPEC}},
  title   = {{SpecJVM2008}},
  note    = {\url{https://www.spec.org/jvm2008/}},
  urldate = {2024-01-31},
  year    = {1998},
}

@Misc{jdk23biconsumer,
  author  = {{Oracle}},
  title   = {{BiConsumer (Java SE 23; JDK 23)}},
  note    = {\url{https://docs.oracle.com/en/java/javase/23/docs/api/java.base/java/util/function/BiConsumer.html}},
  urldate = {2024-01-31},
  year    = {2024},
}

@Misc{jdk23consumer,
  author  = {{Oracle}},
  title   = {{Consumer (Java SE 23; JDK 23)}},
  note    = {\url{https://docs.oracle.com/en/java/javase/23/docs/api/java.base/java/util/function/Consumer.html}},
  urldate = {2024-01-31},
  year    = {2024},
}

@Misc{jdk23streampackage_order,
  author  = {{Oracle}},
  title   = {{java.util.stream (Java SE 23; JDK 23)}},
  note    = {\url{https://docs.oracle.com/en/java/javase/23/docs/api/java.base/java/util/stream/package-summary.html\#Ordering}},
  urldate = {2024-01-31},
  year    = {2024},
}

@misc{jmh,
	author = {{Michael Duigou}},
	title = {{Java Microbenchmarking Harness}},
	year = {2022},
	howpublished = {\url{http://openjdk.java.net/projects/code-tools/jmh/}}
}

@misc{design-complexity,
	author = {{McCabe}},
	title = {{McCabe IQ - Software Metrics Glossary}},
	year = {2022},
	howpublished = {\url{http://www.mccabe.com/iq_research_metrics.htm}}
}

@misc{intellij-idea,
	author = {{JetBrains}},
	title = {{IntelliJ IDEA – the IDE for Pro Java and Kotlin Development}},
	year = {2022},
	howpublished = {\url{https://www.jetbrains.com/idea/}}
}

@misc{oracle-ergonomics,
	author = {{Oracle}},
	title = {{Ergonomics}},
	year = {2022},
	howpublished = {\url{https://docs.oracle.com/javase/8/docs/technotes/guides/vm/gctuning/ergonomics.html}}
}

@misc{oracle-stream,
	author = {{Oracle}},
	title = {{Processing Data with  Java SE 8 Streams, Part 1}},
	year = {2022},
	howpublished = {\url{https://www.oracle.com/technical-resources/articles/java/ma14-java-se-8-streams.html}}
}

@misc{oracle-graalvm,
	author = {{Oracle}},
	title = {{GraalVM}},
	year = {2022},
	howpublished = {\url{https://www.graalvm.org/}}
}

@misc{oracle-jdk,
	author = {{Oracle}},
	title = {{Java Software | Oracle}},
	year = {2022},
	howpublished = {\url{https://www.oracle.com/java/}}
}

@misc{javadoc-distinct,
	author = {{Oracle}},
	title = {{Stream (JDK 24) - distinct}},
	year = {2022},
	howpublished = {\url{https://docs.oracle.com/en/java/javase/24/docs/api/java.base/java/util/stream/Stream.html\#distinct()}}
}

@misc{s2s-code,
	author = {{Filippo Schiavio}},
	title = {{S2S - SQL To Stream}},
	year = {2025},
	howpublished = {\url{http://github.com/usi-dag/S2S}}
}

@misc{jedi,
	author = {{Filippo Schiavio}},
	title = {{JEDI - Java Evaluation of Declarative vs Imperative queries}},
	year = {2025},
	howpublished = {\url{http://github.com/usi-dag/JEDI}}
}

\end{document}